\documentclass[a4paper,twocolumn]{article}
\usepackage[a4paper,left=1.8cm,right=1.8cm,top=2.2cm,bottom=2.2cm]{geometry}
\usepackage{tikz}
\usetikzlibrary{decorations.pathreplacing}
\usepackage{tabularx}
\usepackage{multirow}
\usepackage{amsmath}
\usepackage{amsfonts}
\usepackage{amssymb}
\usepackage{bm}
\usepackage{xcolor}
\usepackage{hyperref}
\hypersetup{colorlinks=true,linkcolor=blue,urlcolor=blue,citecolor=blue}
\usepackage{graphicx}
\usepackage{dcolumn}
\usepackage{mdframed}
\usepackage{etoolbox}
\usepackage{booktabs}
\usepackage{array}
\usepackage{ragged2e}

\usepackage[switch]{lineno}

\newcommand*\patchAmsMathEnvironmentForLineno[1]{%
  \expandafter\let\csname old#1\expandafter\endcsname\csname #1\endcsname
  \expandafter\let\csname oldend#1\expandafter\endcsname\csname end#1\endcsname
  \renewenvironment{#1}%
    {\linenomath\csname old#1\endcsname}%
    {\csname oldend#1\endcsname\endlinenomath}}
\newcommand*\patchBothAmsMathEnvironmentsForLineno[1]{%
  \patchAmsMathEnvironmentForLineno{#1}%
  \patchAmsMathEnvironmentForLineno{#1*}}
\AtBeginDocument{%
  \patchBothAmsMathEnvironmentsForLineno{equation}%
  \patchBothAmsMathEnvironmentsForLineno{align}%
  \patchBothAmsMathEnvironmentsForLineno{flalign}%
  \patchBothAmsMathEnvironmentsForLineno{alignat}%
  \patchBothAmsMathEnvironmentsForLineno{gather}%
  \patchBothAmsMathEnvironmentsForLineno{multline}%
}

\usepackage[T1]{fontenc}

\usepackage{setspace}

\usepackage{achemso}

\usepackage{graphicx}
\usepackage{float}
\newfloat{scheme}{htbp}{los}
\floatname{scheme}{Scheme}
\floatname{chart}{Chart}
\newfloat{graph}{htbp}{loh}

\usepackage{chemformula} % Formulas using \ch{}
\usepackage[version = 4]{mhchem} % Formulas using \ce{}

\newcommand{\trans}{{\mathsf{T}}}
\newcommand{\Gc}{G_{\mathrm{c}}}

\newcommand{\tref}{\tau_{\mathrm{ref}}}
\newcommand{\Tref}{T_{\mathrm{ref}}}
\newcommand{\omegar}{\omega_{\mathrm{r}}}

\usepackage{authblk}
\author[1,2,*]{Isaac Y. Miranda-Valdez}
\author[1]{Kourosh Mobredi}
\author[1]{Juha Koivisto}
\author[1,*]{Mikko J. Alava}
\affil[1]{Department of Applied Physics, School of Science, Aalto University, P.O.\ Box 15600, Espoo, FI-00076 Aalto, Finland}
\affil[2]{Deltan Labs, P.O.\ Box 11000, Espoo, FI-00076 Aalto, Finland}

\title{Time--temperature superposition from multi-temperature chirp rheometry of thermorheologically simple materials via physics-informed Gaussian processes}
\date{*Email: isaac.miranda.valdez@alumni.aalto.fi, mikko.alava@aalto.fi}

\begin{document}

\maketitle
%\linenumbers

\begin{abstract}
Inferring constitutive laws from transient response data is a central inverse problem in soft condensed-matter physics, particularly when the relevant relaxation spectrum extends beyond the time window of a single experiment. Time--temperature superposition (TTS) constrains this problem for thermorheologically simple materials in the linear viscoelastic regime by allowing temperature to rescale the characteristic relaxation time without changing the underlying shape of the relaxation spectrum. We incorporate this constraint into a physics-informed machine-learning framework for time-resolved mechanical spectroscopy via chirp rheometry. We encode compact fractional viscoelastic models as causal convolutional features derived from the Boltzmann superposition principle, and we jointly fit multi-temperature chirp records using sparse variational Gaussian process regression. The model infers horizontal shift factors, constitutive parameters, and an observational-noise scale directly from time-domain stress--strain signals measured in the linear viscoelastic regime, without first extracting frequency-domain spectra or manually shifting separate curves. For synthetic Fractional Maxwell data, the method recovers the generating constitutive law from a prescribed model library and latent shifts over stress-noise levels from 3\% to 25\%. Applied to a DGEBA-IPDA epoxy dataset spanning the glass-transition regime, the pooled likelihood distinguishes competing fractional rheology models using the Bayesian information criterion. These results recast TTS master-curve construction as probabilistic inference over latent time rescaling and linear viscoelasticity, while also identifying limitations, such as homoscedastic noise and weakly informative spectral regimes.
\end{abstract}

\section*{Keywords}

Time-temperature superposition, chirp rheometry, Gaussian processes, machine learning

\section{Introduction}

Many problems in soft condensed-matter physics require inferring an underlying material description from transient, noisy responses that reveal the relevant dynamics only indirectly. In linear viscoelastic rheology, this inverse problem is particularly acute because complex fluids and soft solids can exhibit relaxation timescales far beyond those accessible in a single experiment~\citep{blair1969elementary, song_holten-andersen_mckinley_2023, ricarte_shanbhag_2024}. A finite measurement window therefore reveals only part of the relaxation spectrum, even when the stress response is sampled densely in time.

For thermorheologically simple materials within the linear viscoelastic regime, time–temperature superposition (TTS) provides an additional physical constraint. Temperature changes the characteristic relaxation timescale while leaving the shape of the relaxation spectrum invariant. Measurements acquired at different temperatures can then be represented by one latent master response after applying a temperature-dependent horizontal shift factor \(a_T(T)\) to the time or frequency axis~\citep{ferry1980viscoelastic}. The resulting master curve extends the experimentally accessible range, but its construction is usually treated as a separate analysis step rather than as part of constitutive inference.

Conventionally, TTS requires a sequence of operations that are both experimental and analytical. Mechanical spectra are acquired across multiple temperatures and frequency (or time) ranges, reduced to estimates of the storage and loss moduli ($G^{\prime}$, $G^{\prime\prime}$), and then aligned by horizontal and, where required, vertical shifts. Conventional frequency sweeps can take tens of minutes to acquire each spectrum when the target relaxation window spans many decades. Chirp rheometry addresses this acquisition bottleneck by compressing mechanical spectroscopy into a few seconds~\citep{geri_keshavarz_divoux_clasen_curtis_mckinley_2018, das_vadillo_perego_mckinley_2026, perego_vadillo_mills_das_mckinleyfrs_2025, waeterloos_mckinley_clasen_2025, HUDSONKERSHAW2024105307, holly_1988, winter_chambon_1986, ghiringhelli_roux_bleses_galliard_caton_2012}. Frameworks such as BOTTS~\citep{sheridan_zauscher_brinson_2024} and TORCh~\citep{vadillo_2026_torch} exploit this speed to acquire multi-temperature datasets suitable for master-curve construction, while open-source packages such as \texttt{hermes-rheo}~\citep{Perego2026} streamline the chirp waveform generation and Fourier analysis of experimental data. 

However, the corresponding analysis remains largely sequential. Each chirp is generally converted into separate Fourier estimates of \(G^{\prime}\) and \(G^{\prime\prime}\), after which the spectra are shifted and interpreted together. This transformation reduces a densely sampled transient record to a smaller set of frequency-domain estimates and can propagate phase-estimation errors into the inferred moduli~\citep{angerman2026instrument}. The problem is particularly severe when the viscous contribution is greater than the elastic one (i.e., a loss factor value \(\tan\delta=G^{\prime\prime}/G^{\prime}\ll1\)), because a small phase error can produce a large relative error in the loss modulus. Treating spectral reduction, curve shifting, and constitutive fitting as independent stages also obscures how uncertainty introduced at one stage influences the next~\citep{gergesova_2011, maiti_2019}.

These limitations motivate a different interpretation of TTS. Rather than viewing master-curve construction as post-processing, we treat it as a latent-variable inverse problem. Multi-temperature chirp responses measured for a thermorheologically simple material share a single constitutive viscoelastic description, while temperature modulates the characteristic timescale via record-specific latent shift factors. The shifts and constitutive parameters can therefore be inferred together from the raw stress--strain histories under a common likelihood. This formulation retains the causal structure of the transient data and allows uncertainty in time rescaling and material parameters to be represented within the same model.

The proposed approach connects two related lines of work while changing the object of inference. \citet{hernandez2017thermorheologically}, \citet{pawlak_denisiewicz_2021}, and \citet{genovese_2022_fractional} formulated multi-temperature calibration as an inverse problem in which fractional-derivative parameters and a Williams--Landel--Ferry (WLF) shift law~\citep{williams_landel_ferry_1955} are fitted simultaneously, but their analyses begin with frequency-domain complex moduli. Conversely, \citet{lennon_mckinley_swan_2023b} used Gaussian process regression to align master curves flexibly without specifying a constitutive viscoelastic description. Here, we combine the statistical framework of Gaussian process regression~\citep{rasmussen_2005,alvarez_2013, raissi_2017} with the physical, compact structure of fractional viscoelasticity by embedding candidate linear viscoelastic constitutive kernels and the TTS constraint directly into the raw time-domain records, rather than applying them after spectral extraction. We then use the Bayesian information criterion (BIC)~\citep{leonard2001bayesian, schwarz_1978, kass_raftery_1995} to select the most parsimonious description from a predefined library of constitutive models.

We extend a physics-informed Gaussian-process (GP) framework previously developed for single-chirp inference~\citep{mirandavaldez2026directinferenceviscoelasticmemory} to multi-temperature chirp rheometry. Candidate Fractional Maxwell and Fractional Zener models~\citep{bagley_torvik_1983, bonfanti_kaplan_charras_kabla_2020, jaishankar_mckinley_2013, miranda-valdez_niinisto_makinen_lejon_koivisto_alava_2025} define causal convolutional features derived from the Boltzmann superposition principle. We train a sparse variational GP~\citep{pmlr-v5-titsias09a, hensman2013gaussian} on pooled records from all temperatures, providing a scalable likelihood for shared constitutive parameters, horizontal shift factors, noise scale, and model comparison. We first test parameter and shift recovery on synthetic Fractional Maxwell data with known ground truth. We then analyze a DGEBA-IPDA epoxy dataset spanning the rubbery, glass-transition, and glassy regimes.

\section{Methods}

\begin{figure*}[t]
  \centering
  \includegraphics[width=\textwidth]{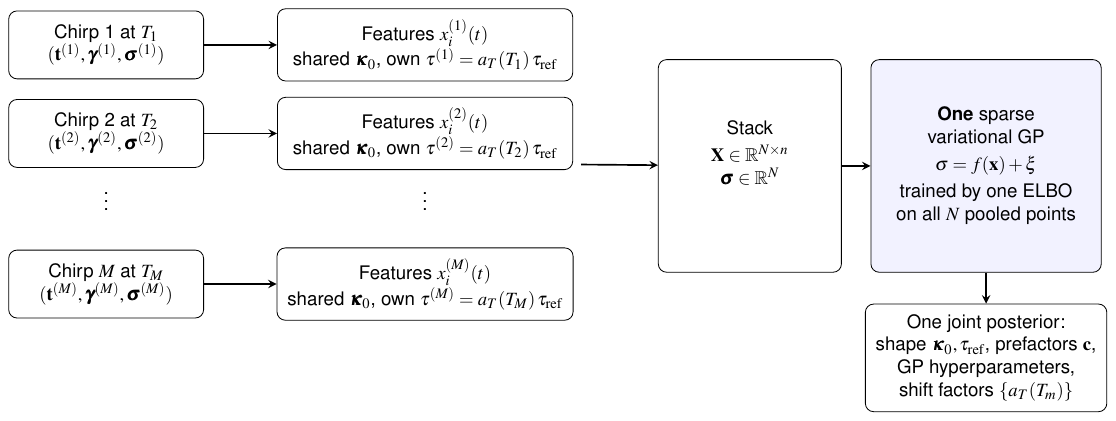}
  \caption{Overview of the joint time-domain inference workflow. Multi-temperature chirp measurements provide the strain $\gamma$ and stress $\sigma$ histories. For each candidate linear viscoelastic model, the measured strain rate $\dot{\gamma}$ is convolved with the constitutive relaxation-kernel components to generate physics-informed features. Temperature-dependent horizontal shift factors $a_T$ rescale the characteristic relaxation time $\tau$, while the kernel shape and constitutive prefactors are shared across records. The features and measured stresses are then pooled and supplied to a sparse variational Gaussian process. Joint inference yields the constitutive parameters, horizontal shift factors, and observational-noise scale, from which the reduced-frequency master curve is constructed. Candidate constitutive models are compared using the Bayesian information criterion.}
  \label{fig:pipeline}
\end{figure*}

The joint inference workflow is summarized in Fig.~\ref{fig:pipeline}. For each temperature, the measured strain rate $\dot{\gamma}$ is first convolved with the relaxation-kernel components of a candidate linear viscoelastic model, producing a set of physics-informed features. The temperature \(T_m\) enters through a latent horizontal shift factor
\(a_T(T_m)\), which rescales the shared reference relaxation time
\(\tref\) to give the record-specific relaxation time
\(\tau^{(m)}=a_T(T_m)\tref\). The dimensionless kernel-shape parameters
and constitutive prefactors remain shared across records. The features and measured stresses are then pooled and used to train a single sparse variational Gaussian process (GP) under a common likelihood. The resulting posterior provides estimates of the constitutive parameters, horizontal shift factors, and observational-noise scale. These estimates are subsequently used to construct a reduced angular frequency $\omegar$ master curve. Repeating the complete inference procedure for each candidate constitutive model enables model comparison using values of BIC.

\subsection{Constitutive physics as a feature map}

Within the linear viscoelastic regime, the stress $\sigma$ response to an imposed strain $\gamma$ history is governed by the Boltzmann superposition principle,
\begin{equation}
  \sigma(t)
  =
  \int_0^t
  G(t-t')\,\dot{\gamma}(t')\,\mathrm{d}t',
  \label{eq:boltzmann}
\end{equation}
where \(t\) is time, \(G(t)\) is the relaxation modulus, and \(\dot{\gamma}(t)\) is the strain rate. Equation~\eqref{eq:boltzmann} expresses the stress as a causal memory of the preceding deformation history. For a given candidate constitutive model, we write the relaxation modulus $G(t)$ as
\begin{equation}
  G(t)
  =
  \sum_{i=1}^{n}
  c_i\,\phi_i(t;\bm{\kappa}),
  \label{eq:kernel_decomp}
\end{equation}
where the kernels \(\phi_i\) encode the assumed viscoelastic topology.
They depend nonlinearly on the kernel parameters \(\bm{\kappa}\), which
may include dimensionless shape parameters, such as fractional exponents,
and characteristic relaxation times. The coefficients \(c_i\), corresponding to moduli or viscosities, enter the constitutive relation linearly.

Substituting Eq.~\eqref{eq:kernel_decomp} into Eq.~\eqref{eq:boltzmann} gives
\begin{equation}
  \sigma(t)
  =
  \sum_{i=1}^{n}
  c_i\,x_i(t;\bm{\kappa}),
  \qquad
  x_i(t;\bm{\kappa})
  =
  \int_0^t
  \phi_i(t-t';\bm{\kappa})
  \dot{\gamma}(t')\,\mathrm{d}t'.
  \label{eq:features}
\end{equation}
Each \(x_i(t)\) is the stress contribution that the \(i\)th kernel component would generate under the measured strain history. Equation~\eqref{eq:features} therefore maps the raw deformation record to a feature space whose coordinates are determined by the candidate constitutive law.

For fixed kernel parameters \(\bm{\kappa}\), which govern the form and
timescale of the relaxation spectrum rather than its overall magnitude,
Eq.~\eqref{eq:features} is linear in the constitutive prefactors \(c_i\). The constitutive model thus determines the feature coordinates, while a regression layer determines how those features combine to reproduce the measured stress. We implement this layer using a GP with a linear covariance function. This choice preserves the linear constitutive structure while providing a probabilistic treatment of the prefactors and observational noise.

Specifically, we place a GP prior on the map from the feature vector
\(\mathbf{x}(t)=(x_1(t),\ldots,x_n(t))^{\trans}\) to the observed stress:
\begin{equation}
  \sigma(t)
  =
  f(\mathbf{x}(t))+\xi(t),
  \qquad
  f\sim
  \mathcal{GP}\!\left[
  0,K(\mathbf{x},\tilde{\mathbf{x}})
  \right],
  \label{eq:gp}
\end{equation}
where \(\xi(t)\) is independent Gaussian observational noise with variance \(s_\xi^2\). The covariance function is
\begin{equation}
  K(\mathbf{x},\tilde{\mathbf{x}})
  =
  s_f^2\,
  \mathbf{x}^{\trans}\tilde{\mathbf{x}},
  \label{eq:linear_kernel}
\end{equation}
where \(s_f^2\) is the signal variance. Because the covariance is linear, the posterior mean remains linear in the physics-informed features. The GP therefore does not replace the constitutive law with an unrestricted nonlinear regression. Instead, it provides a Bayesian regression layer over features generated by that law.

Equations~\eqref{eq:features}--\eqref{eq:linear_kernel} describe the inference problem for a specified relaxation timescale. We next extend this construction to multiple temperatures by allowing each record to have a temperature-dependent timescale while retaining a common relaxation-kernel shape and common constitutive prefactors.

\subsection{Latent time--temperature shifts}

We consider an \(M\) number of chirp records,
\[
\left\{
\left(
\mathbf{t}^{(m)},
\bm{\gamma}^{(m)},
\bm{\sigma}^{(m)}
\right)
\right\}_{m=1}^{M},
\]
measured at temperatures \(T_1,\ldots,T_M\), with \(N_m\) samples in record \(m\). For a thermorheologically simple material, temperature changes the characteristic relaxation time without changing the shape of the linear relaxation spectrum. We therefore define the shared reference-state parameters as
\(\bm{\kappa}_{\mathrm{ref}}=(\bm{\kappa}_0,\tref)\), where
\(\bm{\kappa}_0\) contains the shared dimensionless kernel-shape
parameters and \(\tref\) is the characteristic relaxation time at
the reference temperature \(\Tref\).

The record-specific relaxation time is defined as
\begin{equation}
  \tau^{(m)}
  =
  a_T(T_m)\,\tref,
  \qquad
  a_T(\Tref)=1.
  \label{eq:tts_ansatz}
\end{equation}
The reference condition removes the scale degeneracy between \(\tref\) and the horizontal shift factors. No WLF, Arrhenius, or other parametric temperature law is imposed during training; the independent shift factors are inferred directly from the pooled records.

For record \(m\), the features are constructed using the shared shape parameters \(\bm{\kappa}_0\) and the record-specific relaxation time \(\tau^{(m)}\):
\begin{equation}
  x_i^{(m)}(t)
  =
  \int_0^t
  \phi_i\!\left(
  t-t';
  \bm{\kappa}_0,
  \tau^{(m)}
  \right)
  \dot{\gamma}^{(m)}(t')\,\mathrm{d}t'.
  \label{eq:features_shifted}
\end{equation}
Temperature is therefore not included as a generic GP input. Instead, it enters through a physically constrained latent variable that rescales the memory kernel. Although this formulation could, in principle, be extended to time--concentration or other forms of superposition, we do not explore such extensions in this work.

The record-specific feature matrices
\(\mathbf{X}^{(m)}\in\mathbb{R}^{N_m\times n}\)
and stress vectors
\(\bm{\sigma}^{(m)}\in\mathbb{R}^{N_m}\)
are pooled as
\begin{equation}
  \begin{aligned}
    \mathbf{X} &=
    \begin{bmatrix} \mathbf{X}^{(1)}\\ \vdots\\ \mathbf{X}^{(M)} \end{bmatrix}
    \in\mathbb{R}^{N\times n},
    \qquad
    \bm{\sigma} &=
    \begin{bmatrix} \bm{\sigma}^{(1)}\\ \vdots\\ \bm{\sigma}^{(M)} \end{bmatrix}
    \in\mathbb{R}^{N},
    \\[4pt]
    N &= \sum_{m=1}^{M}N_m .
  \end{aligned}
  \label{eq:stack}
\end{equation}
A single GP is trained on this pooled feature space. The model therefore contains three classes of unknown quantities: shared constitutive parameters, record-specific horizontal shift factors, and GP and variational parameters. We optimize these quantities jointly via Adam~\citep{kingma2017adammethodstochasticoptimization} rather than through separate spectral reduction, curve shifting, and constitutive fitting stages.

Because the pooled datasets can contain hundreds of thousands of time samples, exact GP inference is computationally impractical. We instead maximize a sparse variational evidence lower bound (ELBO) under a Gaussian likelihood~\citep{hensman2013gaussian}. The optimized quantities include \(\bm{\kappa}_0\), \(\tref\), the \(M-1\) independent shift factors, the inducing-point locations, the variational parameters, and the GP hyperparameters. The input features and stress targets are standardized before training. The inducing-point representation reduces the computational cost while retaining a common probabilistic objective for parameter estimation and model comparison. Because the covariance is linear, the required number of inducing points is on the order of the number of input features.

\subsection{Parameter extraction and master-curve construction}

After training, the GP posterior is converted into physically interpretable constitutive parameters. For the linear covariance in Eq.~\eqref{eq:linear_kernel}, the constitutive prefactor associated with feature \(x_i\) is obtained from the posterior sensitivity,
\begin{equation}
  c_i
  =
  \frac{\partial\bar{\sigma}}{\partial x_i},
  \label{eq:prefactor_extraction}
\end{equation}
where \(\bar{\sigma}\) is the posterior-mean stress. After reversing the feature and target standardization, this derivative equals the corresponding posterior weight in physical units. Credible intervals are obtained by applying the same transformation to samples from the posterior distribution of the weights.

The inferred constitutive parameters define an analytic complex modulus for each candidate model. The response at temperature \(T\) is placed on the master curve using the reduced angular frequency
\begin{equation}
  \omegar
  =
  a_T(T)\,\omega.
  \label{eq:reduced_freq}
\end{equation}
An empirical WLF, Arrhenius, or other temperature law may subsequently be fitted to the inferred shift factors, but such a law does not constrain the time-domain likelihood used here.

\begin{figure*}[h]
  \centering
  \includegraphics[width=1\linewidth]{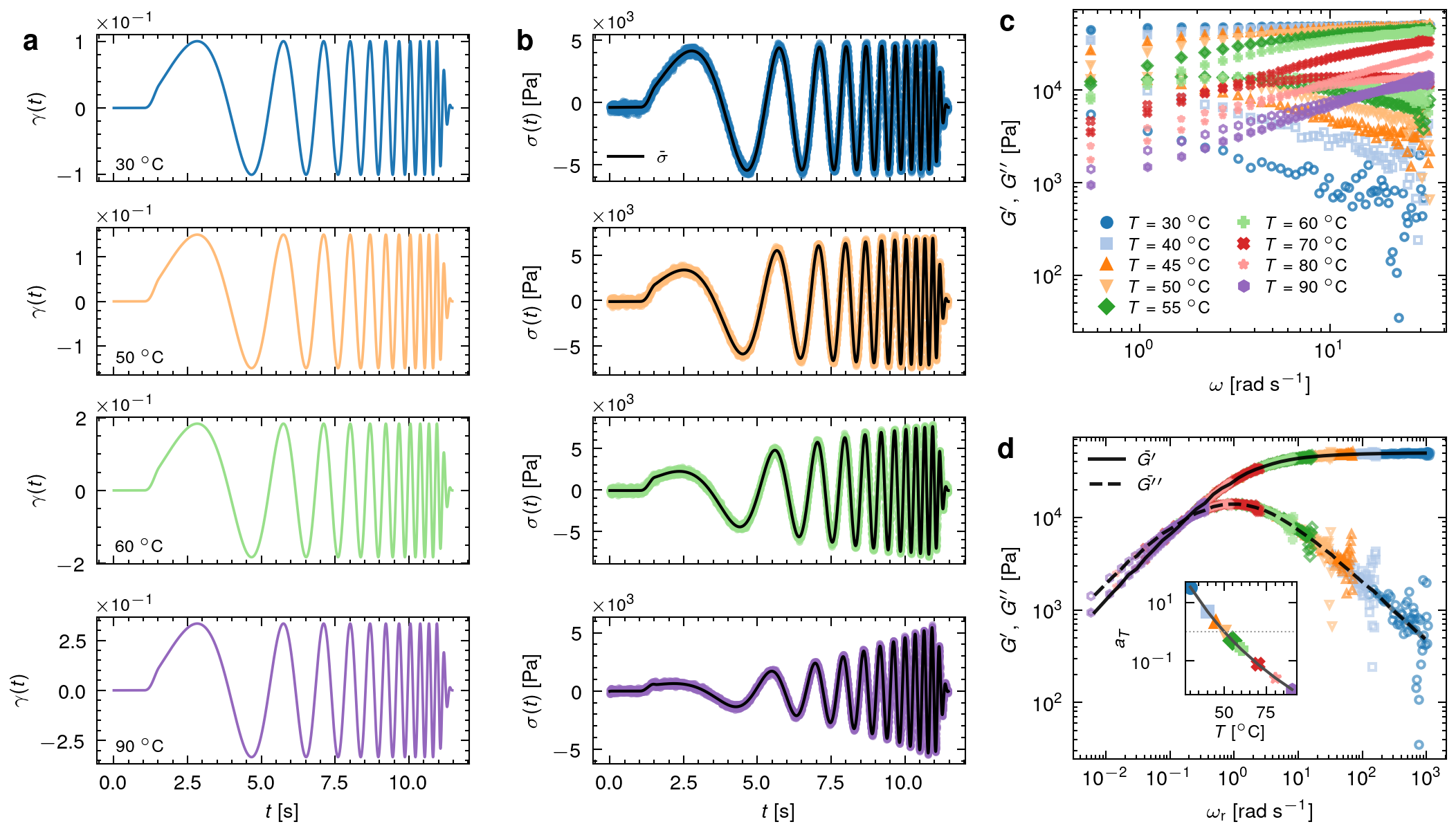}
  \caption{Joint fit of nine synthetic FMM chirps spanning
  $30$--$90~^{\circ}\mathrm{C}$ at $T_{\mathrm{ref}}=50~^{\circ}\mathrm{C}$.
  \textbf{(a)} Imposed chirp strain at four representative temperatures.
  \textbf{(b)} Stress \(\sigma(t)\) (points), posterior-mean prediction \(\bar{\sigma}(t)\) of the joint fit (black), and 95\% credible band (gray), which are imperceptible in this plot because their width is thinner than the line thickness.
  \textbf{(c)} Storage \(G^\prime\) and loss \(G^{\prime\prime}\) moduli obtained by applying the discrete Fourier transform (DFT) separately to each record and plotted against the angular frequency \(\omega\). The low-temperature scatter in \(G^{\prime\prime}\) reflects the ill-conditioning of phase-sensitive spectral estimation when \(\tan\delta\ll1\), rather than a higher imposed noise level.
  \textbf{(d)} Master curve at $\Tref=50~^{\circ}\mathrm{C}$ plotted against the reduced frequency \(\omegar=a_T\omega\), using the shift factors inferred by the joint fit. Solid and dashed black lines are obtained by applying the same DFT reduction to the GP posterior-mean predictions and are not fitted directly to the spectral points.
  \textbf{Inset:} recovered \(a_T\), the reference constraint \(a_T(\Tref)=1\), and the generating WLF law.}
  \label{fig:fmm}
\end{figure*}

\subsection{Candidate models and model comparison}

We define a library composed of constitutive models before fitting the GP to the chirp data. For the synthetic benchmark, the candidates are the Fractional Maxwell model (FMM), Fractional Maxwell Gel (FMG), Fractional Maxwell Liquid (FML), and classical Maxwell model~\citep{jaishankar_mckinley_2013}. For the experimental epoxy data, the library is extended to the Fractional Zener family and its nested limits. Each candidate defines a different set of basis kernels \(\phi_i\), shape parameters, and linear prefactors. We repeat the full inference procedure, including optimizing the shift factors, independently for each candidate. Consequently, we do not evaluate a candidate using shifts estimated under a different constitutive model.

The fitted candidates are compared using
\begin{equation}
  \mathrm{BIC}
  =
  2\mathcal{U}+k\ln N,
  \label{eq:bic}
\end{equation}
where \(\mathcal{U}=-\mathrm{ELBO}\) is used as a surrogate for the negative log marginal likelihood~\citep{pmlr-v5-titsias09a,hensman2013gaussian, kass_raftery_1995, freund_ewoldt_2015}. The parameter count \(k\) includes the fitted constitutive parameters, linear prefactors, \(M-1\) independent horizontal shift factors, and GP hyperparameters. Lower BIC values are preferred.

Because the likelihood is represented by a variational lower bound and the transient observations may be temporally correlated, the resulting BIC values are interpreted as comparative model-selection scores under a common fitting procedure rather than as exact Bayes factors. The sensitivity of the rankings to an autocorrelation-based effective sample size is examined separately.

\section{Results and discussion}

\subsection{Synthetic Fractional Maxwell benchmark}

We first evaluate whether the proposed framework can recover a shared constitutive law, temperature-dependent shift factors, and an observational-noise scale directly from multi-temperature transient data. We also test whether the trained GP predicts the response to an unseen deformation history and whether the pooled likelihood discriminates among competing constitutive models. For this controlled benchmark, we generate synthetic data using a Fractional Maxwell model (FMM).

The FMM consists of two springpots connected in series. A springpot of fractional order \(\alpha\) obeys the constitutive relation
\[
\sigma(t)
=
\mathbb{V}
\frac{\mathrm{d}^{\alpha}\gamma(t)}{\mathrm{d}t^{\alpha}},
\]
where \(\mathrm{d}^{\alpha}\gamma/\mathrm{d}t^{\alpha}\) denotes the fractional derivative of order \(\alpha\), and \(\mathbb{V}\) is the corresponding quasi-property, with units \(\mathrm{Pa\,s^\alpha}\). The limiting cases \(\alpha=0\) and \(\alpha=1\) recover a Hookean spring and a Newtonian dashpot, respectively, whereas \(0\le\alpha\le1\) describes an intermediate power-law element.

For the two springpots in the FMM, we denote the fractional exponents by \(\alpha\) and \(\beta\), with
\(
0\leq\beta<\alpha\leq1,
\)
and their quasi-properties by \(\mathbb{V}\) and \(\mathbb{G}\), with units \(\mathrm{Pa\,s^\alpha}\) and  \(\mathrm{Pa\,s^\beta}\), respectively. Because the elements are connected in series, they experience the same stress, giving the constituve equation
\begin{equation}
  \sigma(t)
  +
  \frac{\mathbb{V}}{\mathbb{G}}
  \frac{\mathrm{d}^{\alpha-\beta}\sigma(t)}
       {\mathrm{d}t^{\alpha-\beta}}
  =
  \mathbb{V}
  \frac{\mathrm{d}^{\alpha}\gamma(t)}
       {\mathrm{d}t^{\alpha}},
  \label{eq:fmm_constitutive}
\end{equation}

In our framework, it is convenient to reparameterize the two quasi-properties in terms of a characteristic relaxation time \(\tref\)  at the reference temperature state and a characteristic modulus \(\Gc\), defined by
\[
\tref
=
\left(
\frac{\mathbb{V}}{\mathbb{G}}
\right)^{1/(\alpha-\beta)},
\qquad
\Gc
=
\mathbb{G}\tref^{-\beta}
=
\mathbb{V}\tref^{-\alpha}.
\]
Thus, \(\Gc\) sets the magnitude of the elastic plateau at high frequencies in Pa units, while \(\tref\) sets the crossover timescale between the two fractional regimes. With this parameterization, the FMM contains one kernel component,
\(n=1\), with prefactor \(c_1=\Gc\). Its dimensionless shape
parameters are \(\bm{\kappa}_0=(\alpha,\beta)\), and its relaxation
time for record \(m\) is
\(\tau^{(m)}=a_T(T_m)\tref\).
 Its relaxation modulus is
\begin{equation}
\begin{aligned}
  G_{\mathrm{FMM}}(t;\tau)
  &=
  \Gc\,
  \phi_{\mathrm{FMM}}(t;\alpha,\beta,\tau),\\
  \phi_{\mathrm{FMM}}(t;\alpha,\beta,\tau)
  &=
  (t/\tau)^{-\beta}
  E_{\alpha-\beta,\,1-\beta}
  \!\left[-(t/\tau)^{\alpha-\beta}\right],
\end{aligned}
\label{eq:fmm_kernel}
\end{equation}
where \(E_{a,b}\) is the two-parameter Mittag--Leffler function~\citep{garrappa_2015}.

The corresponding feature for record \(m\) is
\begin{equation}
\begin{aligned}
  x_{\mathrm{FMM}}^{(m)}(t)
  &=
  \int_0^t
  \phi_{\mathrm{FMM}}\!\left(
  t-t';\alpha,\beta,\tau^{(m)}
  \right)
  \dot{\gamma}^{(m)}(t')\,\mathrm{d}t',\\
  \tau^{(m)}
  &=a_T(T_m)\tref.
\end{aligned}
\label{eq:fmm_feature}
\end{equation}

The \(M-1\)
independent record-specific parameters are the horizontal shift factors
\(a_T(T_m)\), subject to \(a_T(\Tref)=1\). The corresponding
record-specific relaxation times,
\(\tau^{(m)}=a_T(T_m)\tref\), are derived quantities rather than
independently fitted parameters. We generate the synthetic data using
\(\alpha=0.65\), \(\beta=0\),
\(\Gc=5.0\times10^4~\mathrm{Pa}\), and
\(\tref=1.0~\mathrm{s}\) at
\(\Tref=50~^{\circ}\mathrm{C}\). Because the generating value is \(\beta=0\), the synthetic material lies at the Fractional Maxwell Gel (FMG) boundary of the full FMM. This boundary case provides a stringent test of both parameter recovery and model discrimination.

The temperature-dependent shifts follow the WLF relation
\begin{equation}
  \log_{10}a_T(T)
  =
  -\frac{
  C_1\left(T-\Tref\right)
  }{
  C_2+\left(T-\Tref\right)
  },
  \qquad
  a_T(\Tref)=1,
  \label{eq:wlf}
\end{equation}
with \(C_1=8.4\) and \(C_2=130~\mathrm{K}\). We generate nine records at temperatures from \(30\) to \(90~^{\circ}\mathrm{C}\). Each record uses the same \(0.6\)--\(30~\mathrm{rad\,s^{-1}}\) optimally windowed chirp (OWCh)~\citep{geri_keshavarz_divoux_clasen_curtis_mckinley_2018}, sampled at \(500~\mathrm{Hz}\), with 5\% additive stress noise in the baseline case (see Supplementary Material, Sec.~S1). The strain amplitude increases with temperature, as shown in Fig.~\ref{fig:fmm}(a), mimicking experimental protocols in which the excitation amplitude is adjusted to maintain a measurable response across different material regimes.

The joint fit closely reproduces the pooled stress records, with
\(N=51624\),
\(\mathrm{RMSE}\approx166~\mathrm{Pa}\), and
\(R^2\approx0.997\) [Fig.~\ref{fig:fmm}(b)]. Because the synthetic stresses contain independent, zero-mean Gaussian noise, an adequate model should leave residuals that are centered near zero and exhibit little temporal correlation. The record-resolved residuals show no systematic drift (Fig.~S1), and their integrated autocorrelation times remain close to the white-noise limit, \(\tau_{\mathrm{int}}=1\), across most temperatures (Fig.~S2). These diagnostics indicate that the shared constitutive kernel and latent shifts capture the deterministic response without leaving substantial time-correlated structure in the residuals.

The sparse variational GP returns
\(\alpha\approx0.650\) and
\(\Gc\approx4.99\times10^4~\mathrm{Pa}\), with a 95\% credible interval of
\([4.987,4.995]\times10^4~\mathrm{Pa}\) for \(\Gc\). The inferred reference relaxation time differs by approximately 1\% from the generating value \(\tref=1.0~\mathrm{s}\), and the fitted \(\beta\) remains indistinguishable from its boundary value of zero. We repeat the inference at stress-noise levels from 3\% to 25\% to determine whether the parameter recovery persists as the observations become less informative. Across this range, \(\alpha\) remains within 0.3\% and \(\Gc\) within 1.7\% of their generating values (Fig.~S3). The complete fits at 3\%, 8\%, 15\%, and 25\% noise are shown in Figs.~S4--S7. The inferred observational-noise scale also tracks the injected noise in both absolute and relative units (Fig.~S8). Thus, the likelihood separates the dominant constitutive response from the observational noise over the tested range.

Although the model is trained entirely in the time domain, its inferred response must also remain consistent with the corresponding frequency-domain representation via Fourier analysis. Fig.~\ref{fig:fmm}(c) shows \(G^{\prime}\) and \(G^{\prime\prime}\) obtained by applying the discrete Fourier transform (DFT) separately to the observed, noisy records of the stress and plotting the resulting estimates against the unshifted frequency. After applying the GP-inferred shift factors, the individual spectra collapse onto a common reduced-frequency response [Fig.~\ref{fig:fmm}(d)]. The solid and dashed black curves are obtained by applying the same DFT reduction to the posterior-mean stress predictions. Their agreement with the shifted raw DFT points shows that the time-domain model reproduces both the frequency-domain response and its superposition. The inferred shift factors also agree closely with the generating WLF values and with independent estimates obtained using a frequency-domain master-curve method~\citep{lennon_mckinley_swan_2023}, apart from a modest discrepancy at the lowest temperature (Fig.~S9). The low-temperature discrepancy identifies an important limitation of shift-factor inference that we discuss further in the analysis of the epoxy dataset.

\begin{table*}[t]
  \centering
  \caption{BIC-based model comparison for the synthetic Fractional Maxwell benchmark. \(2\mathcal{U}\) and BIC are absolute values for the baseline 5\% noise case. The remaining columns report
  \(\Delta\mathrm{BIC}=\mathrm{BIC}-\mathrm{BIC}_{\min}\)
  at each noise level, such that \(\Delta\mathrm{BIC}=0\) identifies the selected model. All candidates are fitted to the same pooled dataset
  (\(N=51624\)) at each noise level. Lower values are preferred.}
  \label{tab:fmm_model_comparison}
  \small
  \setlength{\tabcolsep}{5pt}
  \renewcommand{\arraystretch}{1.3}
  \begin{tabular*}{\textwidth}{@{\extracolsep{\fill}}llcccccccc}
    \hline
    Model & Circuit & $k$ & $2\mathcal{U}$ & BIC &
    \multicolumn{5}{c}{$\Delta\mathrm{BIC}$} \\
    \cline{6-10}
    & & & 5\% & 5\% & 3\% & 5\% & 8\% & 15\% & 25\% \\
    \hline
    FMG
    & \raisebox{-0.5\height}{\includegraphics[height=0.4in]{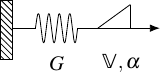}}
    & 15 & $-160749.4$ & $-160586.6$
    & 0.0 & 0.0 & 0.0 & 0.0 & 0.0 \\
    FMM
    & \raisebox{-0.5\height}{\includegraphics[height=0.44in]{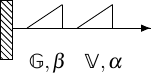}}
    & 16 & $-160752.5$ & $-160578.9$
    & 7.0 & 7.8 & 6.6 & 8.1 & 10.4 \\
    FML
    & \raisebox{-0.5\height}{\includegraphics[height=0.42in]{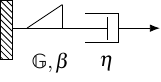}}
    & 15 & $-59151.9$ & $-58989.1$
    & $1.46\times10^{5}$ & $1.02\times10^{5}$
    & $6.37\times10^{4}$ & $2.82\times10^{4}$ & $1.18\times10^{4}$ \\
    Maxwell
    & \raisebox{-0.5\height}{\includegraphics[height=0.4in]{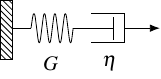}}
    & 14 & $-12100.0$ & $-11948.1$
    & $1.96\times10^{5}$ & $1.49\times10^{5}$
    & $1.05\times10^{5}$ & $5.61\times10^{4}$ & $2.78\times10^{4}$ \\
    \hline
  \end{tabular*}
\end{table*}

We next evaluate the predictive capability of the trained GP using a broadband chirp strain history that is not included in the training data (Fig.~S10). The master curve constructed from the nine training chirps spans
\(5.77\times10^{-3}\) to
\(1.11\times10^{3}~\mathrm{rad\,s^{-1}}\), corresponding to \(5.3\) decades. Using the same sampling interval as the training records,
\(\mathrm{d}t=2\times10^{-3}~\mathrm{s}\) (i.e., 500~Hz), we generate a test chirp six times longer than an individual training chirp
(\(N=34916\) samples). This single chirp spans
\(0.09\) to \(25.7~\mathrm{rad\,s^{-1}}\), or \(2.5\) decades, and therefore covers \(46\%\) of the logarithmic frequency range obtained by superposing the nine training chirps. From the imposed test strain, we compute the physics-informed features using the constitutive parameters inferred from the joint multi-temperature fit and provide them to the trained GP. The GP then accurately predicts the corresponding stress response without refitting (Fig.~S10), acting as a surrogate of the learned constitutive stress--strain relationship.

\subsection{Model discrimination in the synthetic benchmark}

We compare the BIC values obtained by independently fitting the FMM, FMG, Fractional Maxwell Liquid (FML), and classical Maxwell models to the synthetic data generated at different noise levels. The FMG fixes \(\beta=0\) and thus corresponds to the generating model, whereas the full FMM treats \(\beta\) as an additional fitted parameter. Therefore, we aim to test whether the BIC defined in Eq.~\ref{eq:bic} identifies the FMG model as the most parsimonious description, and we evaluate the effect of increasing the noise level in the selection.

Table~\ref{tab:fmm_model_comparison} separates two qualitatively different model-selection regimes. The FML and classical Maxwell models are rejected at every tested noise level by
\(\Delta\mathrm{BIC}\) values of approximately \(10^4\)--\(10^5\) owing to model misfit. The distinction between the FMM and FMG is more subtle because the models are nested and the generating value \(\beta=0\) lies at the boundary of the full FMM. At every tested noise level, the FMG has the lowest BIC, with the full FMM trailing by
\(\Delta\mathrm{BIC}=6.6\)--\(10.4\). In these cases, the improvement in the surrogate likelihood from fitting the additional exponent \(\beta\) does not offset the corresponding complexity penalty, and the BIC favors the most parsimonious description.

Lastly, we investigate the effect of residual autocorrelation. Because the synthetic observations contain independent white noise by construction, an effective-sample-size correction to the BIC in Eq.~\ref{eq:bic} is not required in this benchmark. We nevertheless recompute the comparison using an autocorrelation-based effective sample size, both in the complexity penalty alone and in the likelihood and penalty terms together, as a robustness check. The selected model remains unchanged at every tested noise level (Fig.~S11). We further examine the practical relevance of this correction for the epoxy data, whose residuals exhibit temporal correlation.

\subsection{Experimental DGEBA-IPDA epoxy}

\begin{figure*}[!h]
  \centering
  \includegraphics[width=1\linewidth]{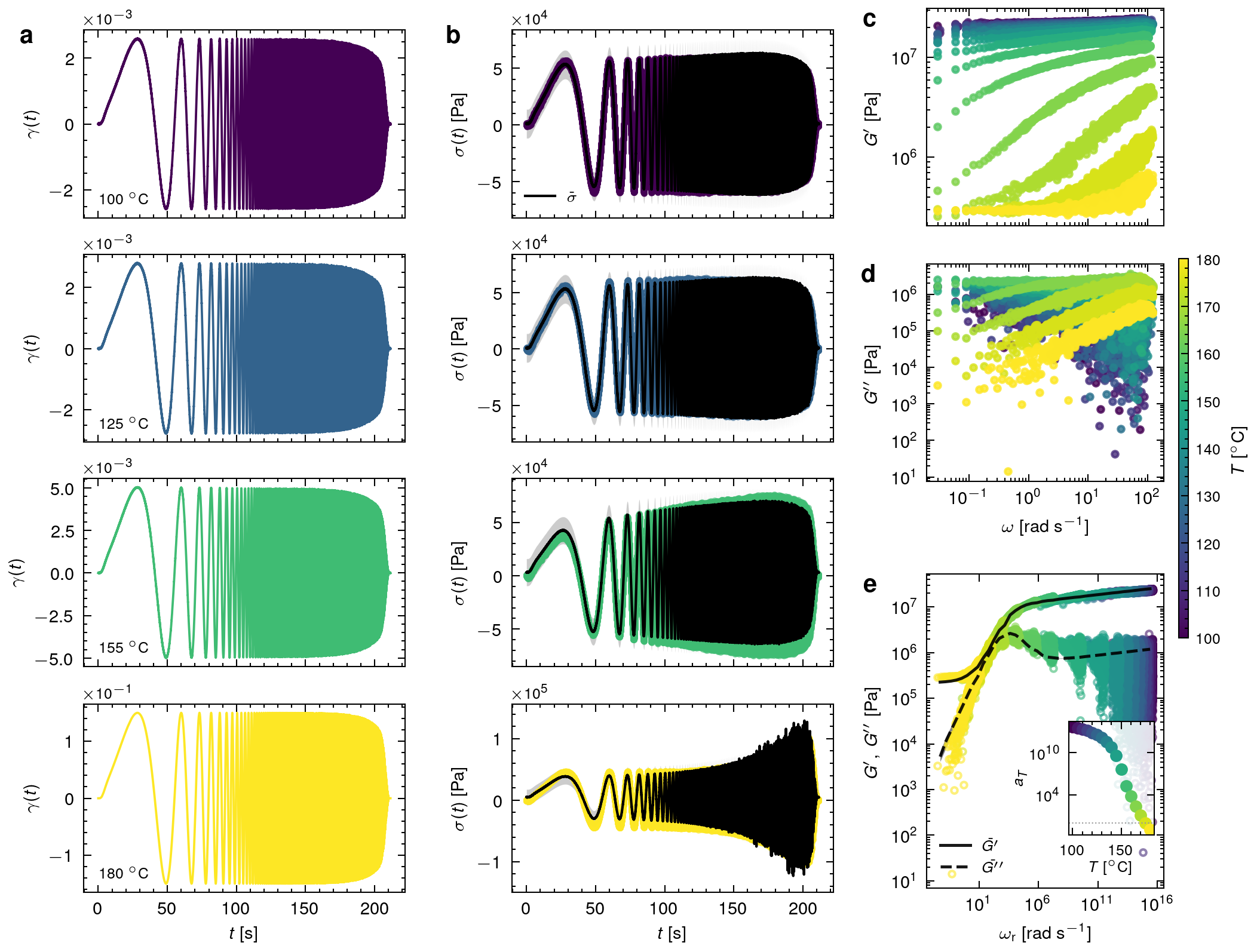}
  \caption{Joint Fractional Zener fit of the DGEBA-IPDA chirp dataset over
  \(100~^{\circ}\mathrm{C}\leq T\leq180~^{\circ}\mathrm{C}\) at $T_{\mathrm{ref}}=175~^{\circ}\mathrm{C}$. Color denotes temperature using one sequential scale.
  \textbf{(a)} Imposed strain at four representative temperatures. The amplitude increases by approximately two orders of magnitude at \(180~^{\circ}\mathrm{C}\), together with a change in axis multiplier, to maintain a measurable torque after softening.
  \textbf{(b)} Measured stress \(\sigma(t)\) (colored points), posterior-mean prediction \(\bar{\sigma}(t)\) (black), and 95\% credible band (gray).
  \textbf{(c,d)} Storage \(G^{\prime}\) and loss \(G^{\prime\prime}\) moduli obtained by applying the DFT separately to each record and plotted against the unshifted angular frequency \(\omega\). The lowest-temperature records lie largely on the glassy plateau, whereas higher-temperature records sample the glass transition.
  \textbf{(e)} Reduced-frequency master curve using \(\omegar=a_T\omega\) at $T_{\mathrm{ref}}=175~^{\circ}\mathrm{C}$. Solid and dashed black lines result from applying the same DFT procedure to the posterior-mean time-domain predictions rather than fitting the spectral points directly.
  \textbf{Inset:} inferred horizontal shift factors; horizontal gray line demarcates the reference temperature state.}
  \label{fig:fzm}
\end{figure*}

We next analyze the multi-temperature chirp dataset for a DGEBA-IPDA epoxy reported by \citet{sheridan_zauscher_brinson_2024}. The records span $100\text{--}180~^{\circ}\mathrm{C}$ in $5~^{\circ}\mathrm{C}$ increments and contain $N=394138$ pooled time samples across the glass-transition regime. Figure~\ref{fig:fzm}(a) presents a representative subset of the chirp records $\gamma(t)$ for different temperatures. The authors assembled the complex modulus $G^{\ast}$ isothermal linear viscoelastic spectra into a master curve and represented it using a Prony-series generalized Maxwell model. The resulting DGEBA-IPDA master curve exhibited the broad glass transition expected for a heterogeneous epoxy thermoset. These observations motivate a constitutive description that captures both a broad distribution of relaxation times and a low-frequency elastic contribution.

\citet{sheridan_zauscher_brinson_2024} noted that the arbitrary strain history used in a broadband chirp obscures the direct connection between the measured stress response and a constitutive representation. With our modeling framework, this connection can be made explicit in the time domain using constitutive convolution features, as we demonstrate above for the case of synthetic data. This avoids an intermediate spectral decomposition during inference and allows the constitutive parameters and horizontal shift factors to be estimated jointly from the original time-domain records.

Here, we replace the discrete Prony-series representation used by
\citet{sheridan_zauscher_brinson_2024} with a compact Fractional Zener model.
The model consists of the FMM branch connected in parallel with an equilibrium
spring of modulus \(G_e\). The spring does not relax, so its shape kernel is
\begin{equation}
  \phi_e(t)=1,
  \label{eq:equilibrium_spring}
\end{equation}
and the corresponding convolutional feature for record \(m\) is
\begin{equation}
  x_e^{(m)}(t)
  =
  \int_0^t
  \phi_e(t-t')\,
  \dot{\gamma}^{(m)}(t')\,\mathrm{d}t'
  =
  \gamma^{(m)}(t)-\gamma^{(m)}(0).
  \label{eq:equilibrium_spring_feature}
\end{equation}
Assuming zero initial strain, \(x_e^{(m)}(t)=\gamma^{(m)}(t)\).

Because the FMM branch and the equilibrium spring are connected in parallel,
their relaxation moduli add. Therefore, the
Fractional Zener relaxation modulus can be written as
\begin{equation}
  G_{\mathrm{FZM}}(t)
  =
  G_{\mathrm{FMM}}(t)+G_e.
  \label{eq:fzm}
\end{equation}
The model therefore has \(n=2\) features, shared dimensionless
shape parameters \(\bm{\kappa}_0=(\alpha,\beta)\), a shared
reference relaxation time \(\tref\), record-specific relaxation
times \(\tau^{(m)}=a_T(T_m)\tref\), and linear prefactors
\(\mathbf{c}=(G_c,G_e)\).

The joint model reproduces the stress histories over the full temperature range, with
$\mathrm{RMSE}=6158~\mathrm{Pa}$ and
$R^2=0.982$ (Fig.~\ref{fig:fzm}b). We verified model convergence by examining the optimization traces of the shape parameters, which approach stable values (Fig.~S12). The DFT-derived spectra show the expected temperature separation before shifting (Fig.~\ref{fig:fzm}c,d). The low-temperature records lie mainly on the glassy plateau, while for higher temperatures records traverse the softer transition regime. Applying the inferred horizontal shifts produces a reduced-frequency master response spanning approximately 14.2 decades (Fig.~\ref{fig:fzm}e). The rapid decrease of \(a_T\) with temperature seen in the inset of Fig.~\ref{fig:fzm}e is qualitatively consistent with WLF-type behavior near the glass transition.

The fitted value \(\beta=0.030\) lies close to the elastic boundary \(\beta=0\), raising the question of whether the additional flexibility of the general FZM is warranted or whether a more parsimonious model can describe the pooled data adequately. At the same time, the shape of the loss modulus in Fig.~\ref{fig:fzm}(e) provides a physical reason for retaining \(\beta\). The two fractional exponents allow the model to represent different asymptotic behaviors at short and long times, or equivalently at high and low reduced frequencies, and thereby capture the asymmetric shape of \(G^{\prime\prime}\) across the relaxation regime. Fixing \(\beta=0\) constrains one of these asymptotic branches and gives a less adequate representation of the observed transition.

To determine whether this additional flexibility is supported
statistically, we fit each candidate model independently. The tested library compares the results from Fig.~\ref{fig:fzm} with the FZM with \(\beta\) fixed at zero, the FMM, the FMG, and the classical Zener model. We then compare the independently optimized fits using the BIC. Table~\ref{tab:epoxy_model_comparison} and Fig.~S13 summarize the fitted shape parameters, relaxation times, prefactors, goodness-of-fit statistics, and model-selection scores.

Under the BIC expression in Eq.~\ref{eq:bic}, the general FZM is preferred, followed by the FMM, the nested FZM with \(\beta=0\), the FMG, and the classical Zener model. This ordering, shown in Table~\ref{tab:epoxy_model_comparison}, is driven primarily by the likelihood term rather than by the complexity penalty. The candidates differ by only one or two fitted parameters, whereas \(2\mathcal{U}\) changes substantially across the models. The same trend is reflected in the goodness-of-fit statistics: \(R^2\) decreases from \(0.9818\) for the general FZM to \(0.7888\) for the classical Zener model. Thus, under the complete sample-size assumption, the improvement in fit provided by the general FZM more than compensates for its additional parameter.

Because the experimental residuals are temporally correlated, the sample size \(N=394138\) overstates the number of independent observations. Recalculating the BIC using the autocorrelation-based effective sample size (Fig.~S14 and Fig.~S15) substantially reduces the differences among the candidate models but leaves the selection of the general FZM unchanged. The corrected ranking places the nested FZM with \(\beta=0\) second, followed by the FMM, FMG, and classical Zener model. The similar $\Delta\mathrm{BIC}$ scores of the nested FZM, FMM, and FMG show that their relative ordering is sensitive to the treatment of residual correlation. Nevertheless, the general FZM remains preferred under both calculations. The nonzero value of \(\beta\) is therefore supported both physically, through the asymmetric shape of \(G^{\prime\prime}\) in Fig.~\ref{fig:fzm}(e), and statistically, through the model-selection scores.

\begin{table*}[t]
  \centering
  \caption{Model comparison for the DGEBA-IPDA chirp dataset over
  \(100\)--\(180~^{\circ}\mathrm{C}\), using \(N=394138\) pooled time
  samples. Here, \(\mathcal{U}=-\mathrm{ELBO}\), and
  \(\Delta\mathrm{BIC}=\mathrm{BIC}-\mathrm{BIC}_{\min}\), with
  \(\mathrm{BIC}=2\mathcal{U}+k\ln N\).
  The quantity \(\Delta\mathrm{BIC}_{\mathrm{eff}}\) is the corresponding
  difference after accounting for the autocorrelation-based effective
  sample size \(N_{\mathrm{eff}}\). Lower values are preferred.
  The reported values of \(2\mathcal{U}\), \(\Delta\mathrm{BIC}\), and
  \(\Delta\mathrm{BIC}_{\mathrm{eff}}\) are scaled by \(10^5\).
  RMSE is reported in Pa, the relaxation time in ms, and the moduli in
  MPa. A dagger denotes a parameter fixed at zero, and a dash denotes a
  parameter that is not applicable.}
  \label{tab:epoxy_model_comparison}

  \small
  \setlength{\tabcolsep}{1.2pt}
  \renewcommand{\arraystretch}{1.30}

  \begin{tabular*}{\textwidth}
    {@{\extracolsep{\fill}}lcrrrrrrrrrcc}
    \hline
    Model
    & \(k\)
    & \(2\mathcal{U}\)
    & \(\Delta\mathrm{BIC}\)
    & \(\Delta\mathrm{BIC}_{\mathrm{eff}}\)
    & \(N_{\mathrm{eff}}\)
    & RMSE
    & \(R^2\)
    & \(\alpha\)
    & \(\beta\)
    & \(\tref\) [ms]
    & \(G_c\) [MPa]
    & \(G_e\) [MPa] \\
    \hline

    \textbf{FZM}
    & 25
    & \(-4.56702\)
    & 0
    & 0
    & 17447.0
    & 6156
    & 0.9818
    & 0.556
    & 0.0298
    & \(4.22\times10^{-1}\)
    & 10.740
    & 0.2311 \\

    FMM
    & 24
    & \(-2.74115\)
    & 1.826
    & 0.188
    & 1936.3
    & 7774
    & 0.9709
    & 0.437
    & 0.0200
    & \(1.10\times10^{-1}\)
    & 14.217
    & -- \\

    FZM (\(\beta=0\))
    & 24
    & \(-2.08187\)
    & 2.485
    & 0.154
    & 9028.0
    & 8452
    & 0.9656
    & 0.491
    & \(0^{\dagger}\)
    & \(1.00\times10^{-1}\)
    & 19.215
    & 0.1912 \\

    FMG
    & 23
    & \(-1.16919\)
    & 3.398
    & 0.191
    & 3530.3
    & 9496
    & 0.9566
    & 0.460
    & \(0^{\dagger}\)
    & \(1.00\times10^{-1}\)
    & 19.359
    & -- \\

    Zener
    & 23
    & \(5.05969\)
    & 9.626
    & 0.336
    & 10413.0
    & \(2.095\times10^4\)
    & 0.7888
    & --
    & --
    & \(9.95\times10^{-1}\)
    & 13.986
    & 0.4695 \\

    \hline
  \end{tabular*}
\end{table*}

Having established the preferred constitutive model, we next examine the inferred horizontal shift factors, shown in the inset of Fig.~\ref{fig:fzm}(e), by comparing them with two frequency-domain estimates. These are the shifts obtained using the data-driven method of \citet{lennon_mckinley_swan_2023b} and those reported by \citet{sheridan_zauscher_brinson_2024}, who applied pairwise horizontal shifting to minimize the noise-weighted mismatch in \(G^{\prime}\) and \(G^{\prime\prime}\). The three estimates agree broadly at high temperatures and through the glass-transition region but diverge at the lowest temperatures (Figs.~S16 and S17), demonstrating that master-curve alignment is not uniquely determined over the full temperature range.

This pattern is physically and statistically informative. Near the transition, the measured response has substantial slope and curvature, and neighboring temperatures sample overlapping but distinguishable portions of the relaxation spectrum. Horizontal displacement is therefore constrained by identifiable features in the data. At low temperatures, by contrast, the measured window lies largely on the glassy plateau. A large horizontal shift can then produce only a small change over the observed frequency interval, making \(a_T\) weakly identifiable. In this regime, the inferred shift depends more strongly on the representation and optimization objective of the superposition method.

Although no estimate can be identified as the uniquely correct shift without independent ground truth, the present method has the advantage of inferring all shift factors jointly from the raw stress--strain records under a common causal constitutive model and likelihood. It therefore avoids phase-sensitive spectral estimation and pairwise curve alignment. This integration also introduces an important limitation: the inference is tied to the prescribed constitutive-model library, and if none of the candidate models adequately represents the material response, no valid solution can be obtained within the framework. Thus, the method provides a more integrated physical and statistical basis for the inferred shifts, but it remains sensitive to constitutive-model misspecification and does not eliminate weak identifiability in featureless plateau regions, such as the low-temperature glassy plateau at high reduced angular frequencies or an equilibrium plateau, when present.

\section{Conclusion}

We formulated time--temperature superposition as a joint probabilistic inverse problem using raw multi-temperature chirp records. Fractional relaxation kernels generate causal, physics-informed convolutional features, and a sparse variational GP jointly infers the constitutive parameters, horizontal shift factors, and observational-noise scale. The approach therefore combines constitutive inference and master-curve construction without prior spectral reduction or manual curve alignment.

For the synthetic FMM benchmark, the method recovers the constitutive law, relaxation scale, noise level, and latent shifts over stress-noise levels from 3\% to 25\%. For the experimental DGEBA-IPDA dataset, the FZM is preferred among the tested models and yields a reduced-frequency response spanning approximately 14.2 decades. Comparisons with frequency-domain methods recover the same dominant shifting trend but reveal weak shift-factor identifiability on the low-temperature glassy plateau. Furthermore, model discrimination is reliable for clearly different constitutive structures and remains sensitive to small likelihood differences for nested fractional models near a parameter boundary.

The present framework is limited to horizontal shifts and assumes homoscedastic, temporally independent Gaussian noise. Because chirps acquired at different temperatures can have different torque amplitudes and signal-to-noise ratios relative to the instrumental resolution, future work should consider a record- or signal-dependent heteroscedastic likelihood. Vertical shifts would require record-specific factors \(b_T(T_m)\), a reference constraint, and additional physical regularization to distinguish them from the shared constitutive prefactors and horizontal shifts. The feature construction also requires relaxation-modulus kernels with a characteristic timescale that can be rescaled by \(a_T\). Models formulated primarily through creep compliance and retardation times, including common fractional Kelvin--Voigt variants, are therefore not directly compatible without deriving an equivalent relaxation kernel or reformulating the inference in terms of creep compliance.

Overall, the framework shows how physics-informed convolutional features can connect machine learning with rheological inverse problems. Constitutive convolution provides a causal, interpretable representation, while probabilistic regression estimates constitutive parameters, temperature-dependent time rescaling, and observational uncertainty. The master curve is thus one outcome of a broader inference that also provides constitutive parameters, model comparisons, and information about where the response is well constrained or remains sensitive to modeling assumptions.

\section*{Supplementary Material}

The Supplementary Material is available free of charge at [link to be inserted after acceptance].

The Supplementary Material contains the additional synthetic-noise benchmarks, recovery of the observational-noise scale, residual and autocorrelation diagnostics, effective-sample-size analyses, optimization histories, parameter-recovery trends, and independent comparisons of the inferred shift factors and reduced-frequency master curves. The file also includes results for two additional datasets of thermorheologically simple materials in Fig.~S18 and S19.

\section{Acknowledgments}
\textbf{Funding:} I.~M.~V. thanks the Vilho, Yrj\"{o}, and Kalle V\"{a}is\"{a}l\"{a} Foundation of the Finnish Academy of Science and Letters and the Finnish Foundation for Technology Promotion for personal funding. The authors acknowledge funding from the FinnCERES flagship (151830423), Business Finland (210129, 211835, 211909, 211989), the Future Makers program, and the Finnish Cultural Foundation. K. M. and M. J. A. acknowledge the financial support provided by the Finnish Ministry of Education and Culture through the Pilot for Doctoral Programmes (Pilot Project: Mathematics of Sensing, Imaging and Modelling). The authors acknowledge funding from the European Union: the ARCHIBIOFOAM project has received funding from the European Union's Horizon Europe research and innovation program under grant agreement No 101161052. Views and opinions expressed are, however, those of the authors only and do not necessarily reflect those of the European Union or the European Innovation Council and SMEs Executive Agency (EISMEA). Neither the European Union nor the granting authority can be held responsible for them. The authors acknowledge the computational resources provided by the Aalto Science-IT project, and thank Mohua Das and Prof.\ Gareth McKinley for their guidance on the numerical simulation of chirps.

\textbf{Author contributions:} I.~M.~V.: Conceptualization, Methodology, Software, Formal analysis, Visualization, Investigation, Writing---original draft, Writing---review and editing. K.~M.: Formal analysis, Validation, Writing---review and editing. J.~K.: Supervision, Validation, Writing---review and editing. M.~A.: Supervision, Validation, Writing---review and editing, Funding acquisition, Project administration.

\textbf{Competing interests:} The authors declare that they have no competing interests.

\textbf{Data and materials availability:} All data needed to evaluate the conclusions in the paper are present in the paper and/or the Supplementary Materials. The data and code that support the findings of this study are openly available in GitHub repository RheoGP (v1.1.0) at https://github.com/mirandi1/rheogp and 10.5281/zenodo.23558075, Ref.~\citep{miranda_valdez_2026_23558075}

\bibliography{article_main}

%%%%%%%%%%%%%%%%%%%%%%%%%%%%%%%%%%%%%%%%%%%%%%%%%%%%%%%%%%%%%%%%%%%%%
%% If you are using classical BibTeX rather than biblatex,
%% remove the \printbibliography and uncomment the \bibliograpy one
%%%%%%%%%%%%%%%%%%%%%%%%%%%%%%%%%%%%%%%%%%%%%%%%%%%%%%%%%%%%%%%%%%%%%
%\printbibliography
%\bibliography{references.bib}

\newpage

%\rule{0.05in}{1.75in}%
\begin{minipage}[b][1.75in]{3.25in}
  \sffamily
  \frenchspacing
\includegraphics[width=1\linewidth]{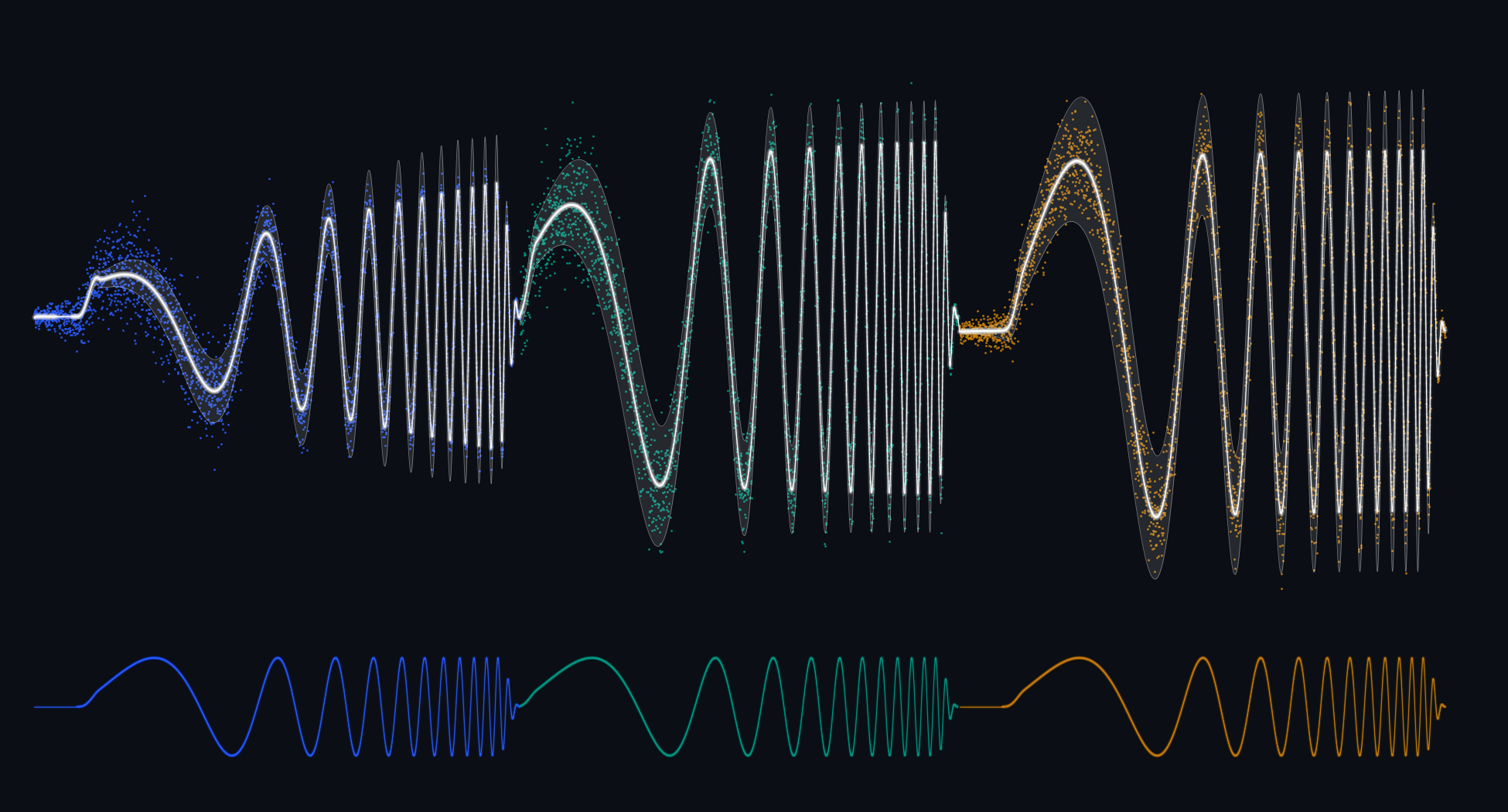}
  
\end{minipage}%
%\rule{0.05in}{1.75in}

\end{document}

\title{Supplementary Information: Time--temperature superposition from multi-temperature chirp rheometry of thermorheologically simple materials via physics-informed Gaussian processes}

\author{Isaac Y. Miranda-Valdez}
\email[]{isaac.miranda.valdez@alumni.aalto.fi}
\affiliation{Department of Applied Physics, School of Science, Aalto University, P.O.\ Box 15600, Espoo, FI-00076 Aalto, Finland}
\affiliation{Deltan Labs, P.O.\ Box 11000, Espoo, FI-00076 Aalto, Finland}

\author{Kourosh Mobredi}
\affiliation{Department of Applied Physics, School of Science, Aalto University, P.O.\ Box 15600, Espoo, FI-00076 Aalto, Finland}

\author{Juha Koivisto}
\affiliation{Department of Applied Physics, School of Science, Aalto University, P.O.\ Box 15600, Espoo, FI-00076 Aalto, Finland}

\author{Mikko J. Alava}
\email[]{mikko.alava@aalto.fi}
\affiliation{Department of Applied Physics, School of Science, Aalto University, P.O.\ Box 15600, Espoo, FI-00076 Aalto, Finland}

\date{\today}
\maketitle

%% ============================================================
\section*{S1.\; Numerical simulation of chirp rheometry}
\label{sec:SI_chirp}
%% ============================================================

We obtained the numerical simulations from the main paper using a strain-controlled, optimally windowed chirp ($\gamma$-OWCh) protocol following Geri \textit{et al.}~\cite{geri_keshavarz_divoux_clasen_curtis_mckinley_2018}. The imposed strain is
\begin{equation}
    \gamma(t) = \gamma_0\,w(t)\,\sin[\psi(t)],
    \label{eq:SI_chirp}
\end{equation}
where $\gamma_0$ is the nominal strain amplitude, $w(t)$ is a dimensionless tapering window, and $\psi(t)$ is the accumulated phase. To probe the frequency interval $[\omega_1,\omega_2]$ over a duration $T_{\mathrm{owc}}$, the chirps use a logarithmic sweep,
\begin{equation}
    \psi(t) = \frac{\omega_1 T_{\mathrm{owc}}}{\ln R}\left[\exp\!\left(\frac{t}{T_{\mathrm{owc}}}\ln R\right) - 1\right] + \psi_0, \qquad R = \frac{\omega_2}{\omega_1},
    \label{eq:SI_log_sweep}
\end{equation}
so that the instantaneous angular frequency $\omega(t) = \mathrm{d}\psi/\mathrm{d}t$ increases exponentially from $\omega_1$ to $\omega_2$; $\psi_0$ is an initial phase offset, zero unless stated otherwise. To reduce spectral leakage, a chirp is tapered with a Tukey-type window,
\begin{equation}
    w(t) =
    \begin{cases}
    \cos^2\!\left[\dfrac{\pi}{r}\!\left(\dfrac{t}{T_{\mathrm{owc}}}-\dfrac{r}{2}\right)\right], & 0\leq \dfrac{t}{T_{\mathrm{owc}}}\leq \dfrac{r}{2},\\[8pt]
    1, & \dfrac{r}{2}<\dfrac{t}{T_{\mathrm{owc}}}<1-\dfrac{r}{2},\\[8pt]
    \cos^2\!\left[\dfrac{\pi}{r}\!\left(\dfrac{t}{T_{\mathrm{owc}}}-1+\dfrac{r}{2}\right)\right], & 1-\dfrac{r}{2}\leq \dfrac{t}{T_{\mathrm{owc}}}\leq 1,
    \end{cases}
    \label{eq:SI_tukey}
\end{equation}
where $r\in(0,1)$ is the taper fraction; the reported data used $r=0.10$ throughout.

%% ============================================================
\section*{S2.\; Complementary results for the numerical simulation}
\label{sec:SI_features}
%% ============================================================

%% ------------------------------------------------------------
\subsection*{S2.1.\; Residual diagnostics}
\label{sec:si_fmm_diagnostics_model_selection}
%% ------------------------------------------------------------

We next examined the residual structure of the synthetic FMM benchmark used in the main text. Because joint inference is performed directly in the time domain, residual diagnostics directly check whether the fitted constitutive memory and inferred time-rescaling variables capture the deterministic stress response, leaving only the imposed observational noise.

\begin{figure*}[t]
    \centering
    \includegraphics[width=1\linewidth]{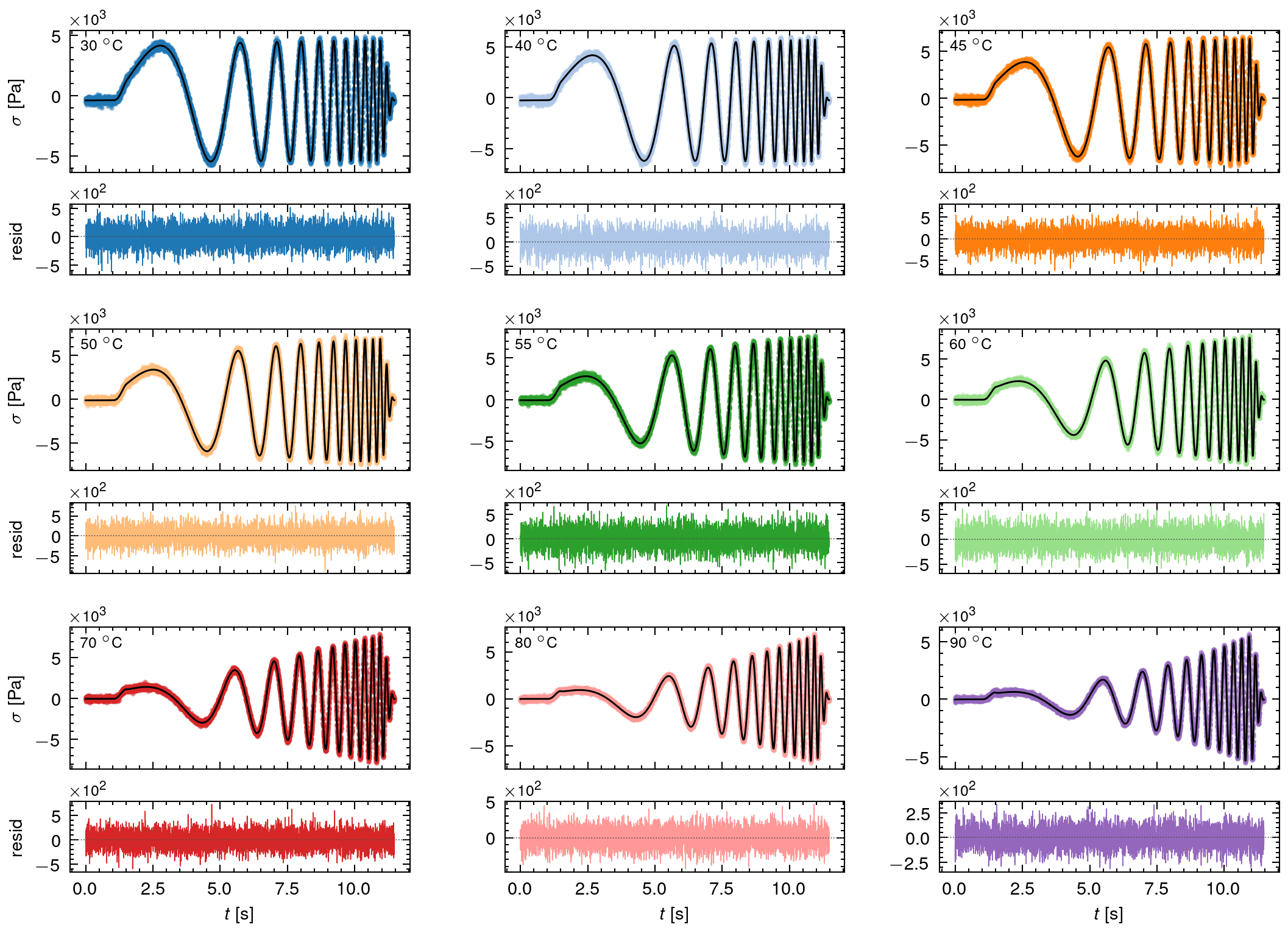}
    \caption{Time-domain residuals for the synthetic FMM benchmark at 5\% additive stress noise. Residuals are shown separately for each temperature-dependent chirp record. The residuals remain centered around zero over the chirp duration, indicating that the inferred constitutive response and latent horizontal shifts capture the deterministic stress response across temperatures.}
    \label{fig:si_fmm_residuals_per_chirp}
\end{figure*}

Figure~\ref{fig:si_fmm_residuals_per_chirp} shows the residuals for the 5\% noise case from the main paper, separated by chirp record at every synthetic temperature. The residuals remain centered around zero across temperatures, with no systematic drift over the chirp duration. This indicates that the joint model captures the dominant temperature-dependent viscoelastic response and that the remaining discrepancies are consistent with observational noise.

\begin{figure*}[t]
    \centering
    \includegraphics[width=1\linewidth]{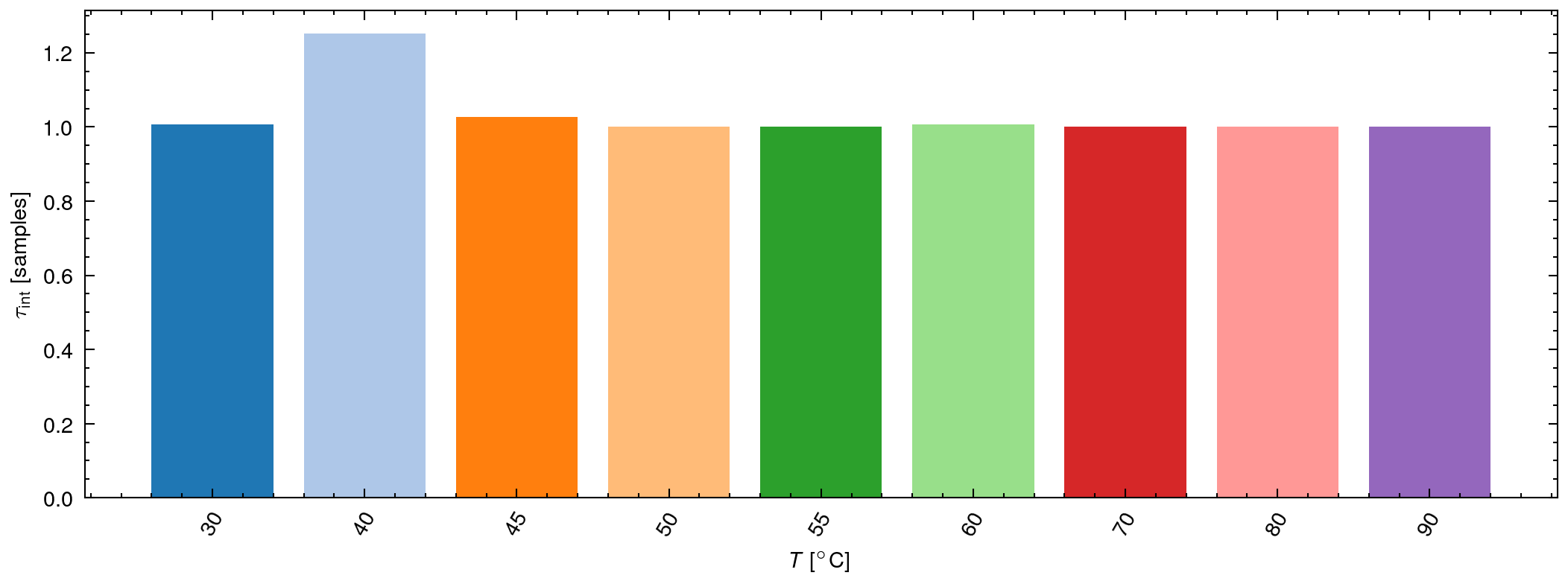}
    \caption{Residual autocorrelation analysis for the synthetic FMM benchmark at 5\% additive stress noise. The bar plot shows the integrated residual autocorrelation time, $\tau_{\mathrm{int}}$, for each temperature-dependent chirp record. These estimates quantify the reduction in independent information caused by temporal correlation in the residuals and are used to define an effective sample size for model-selection diagnostics.}
    \label{fig:si_fmm_residual_autocorr}
\end{figure*}

To quantify possible temporal correlations in the residuals, we computed the residual autocorrelation for each chirp record. Figure~\ref{fig:si_fmm_residual_autocorr} summarizes the resulting autocorrelation time scale, $\tau_{\mathrm{int}}$, as a function of temperature. The estimated autocorrelation time provides a measure of the number of effectively independent samples in each time-domain record and motivates the use of an effective sample size when comparing models by information criteria.

%% ------------------------------------------------------------
\subsection*{S2.2.\; Synthetic FMM benchmark at additional noise levels}
\label{sec:si_fmm_noise_levels}
%% ------------------------------------------------------------

To assess the robustness of the joint time-domain inference procedure, we repeated the synthetic Fractional Maxwell model (FMM) benchmark described in the main text at four additional stress-noise levels. In addition to the baseline 5\% case discussed in the main article, we considered additive Gaussian stress noise with amplitudes of 3\%, 8\%, 15\%, and 25\% of the noiseless stress scale. For each noise level, the same nine-temperature chirp protocol, WLF shift law, reference temperature, and generating FMM parameters were used:
\[
\alpha=0.65,\qquad
\beta=0,\qquad
G_c=5.0\times10^4~\mathrm{Pa},\qquad
\tau_{\mathrm{ref}}=1.0~\mathrm{s},
\]
with $T_{\mathrm{ref}}=50~^{\circ}\mathrm{C}$ and temperatures spanning $30$--$90~^{\circ}\mathrm{C}$. We also kept the imposed chirp frequency range and sampling rate fixed, so changes in the inference results can be attributed directly to increased observational noise.

For every noise level, we trained the same sparse variational Gaussian-process model on the pooled time-domain stress--strain records. We inferred the horizontal shift factors independently for each temperature, subject to the reference constraint $a_T(T_{\mathrm{ref}})=1$, and shared the constitutive parameters across all records. We imposed no WLF form during training; as in the main text, we compared the generating WLF law only after inferring the latent shift factors.

We assessed how the recovered constitutive and time--temperature superposition parameters vary with noise level. Because the BIC variants based on $N$ and $N_{\mathrm{eff}}$ give the same model-selection decision for these synthetic cases (Sec.~S2.5), the comparison is reported for the BIC-winning model obtained using the full-sample criterion.

\begin{figure*}[t]
    \centering
    \includegraphics[width=1\linewidth]{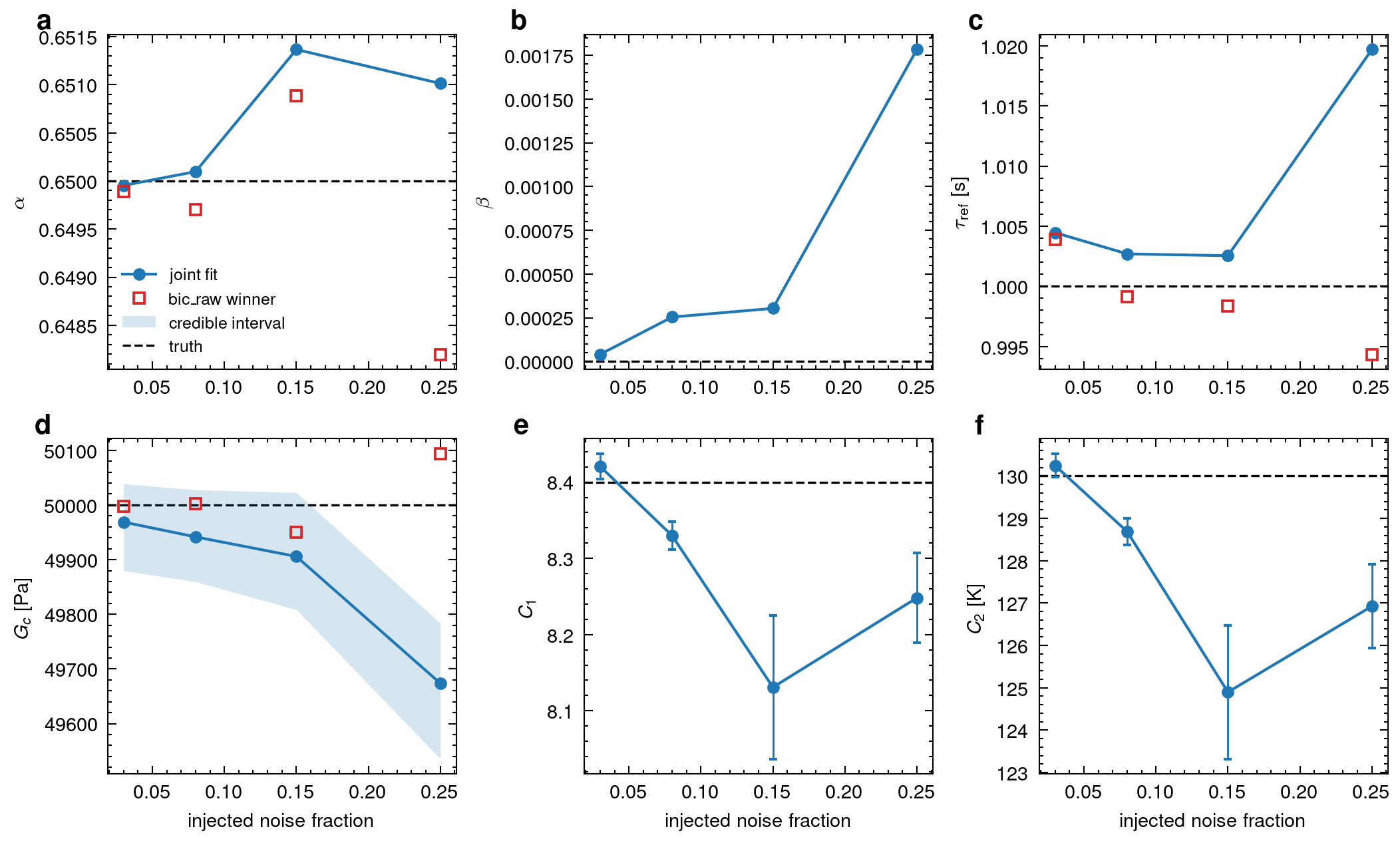}
    \caption{Parameter recovery across synthetic FMM noise levels. Parameters are shown for the generating FMM and for the BIC-winning model selected using the full-sample criterion. \textbf{(a)} Fractional exponent $\alpha$ as a function of injected noise level. \textbf{(b)} Exponent $\beta$. \textbf{(c)} Characteristic time scale at the $\tau_{\mathrm{ref}}$. \textbf{(d)} Modulus scale $G_c$. \textbf{(e)} WLF coefficient $C_1$. \textbf{(f)} WLF coefficient $C_2$. The comparison shows how increasing observational noise affects recovery of the constitutive and time--temperature superposition parameters.}
    \label{fig:si_fmm_noise_parameter_recovery}
\end{figure*}

Figure~\ref{fig:si_fmm_noise_parameter_recovery} compares the recovered parameters as a function of injected noise level for the generating FMM and for the BIC-winning model. Panels~(a)--(d) show the evolution of the constitutive parameters $\alpha$, $\beta$, $\tau_{\mathrm{ref}}$, and $G_c$, respectively. Panels~(e) and~(f) show the recovered WLF coefficients $C_1$ and $C_2$. The parameter estimates remain close to the generating values at low and moderate noise levels, while increasing noise broadens the uncertainty and leads to larger deviations, particularly for parameters associated with weaker or nearly nested model components.

\begin{figure*}[t]
    \centering
    \includegraphics[width=1\linewidth]{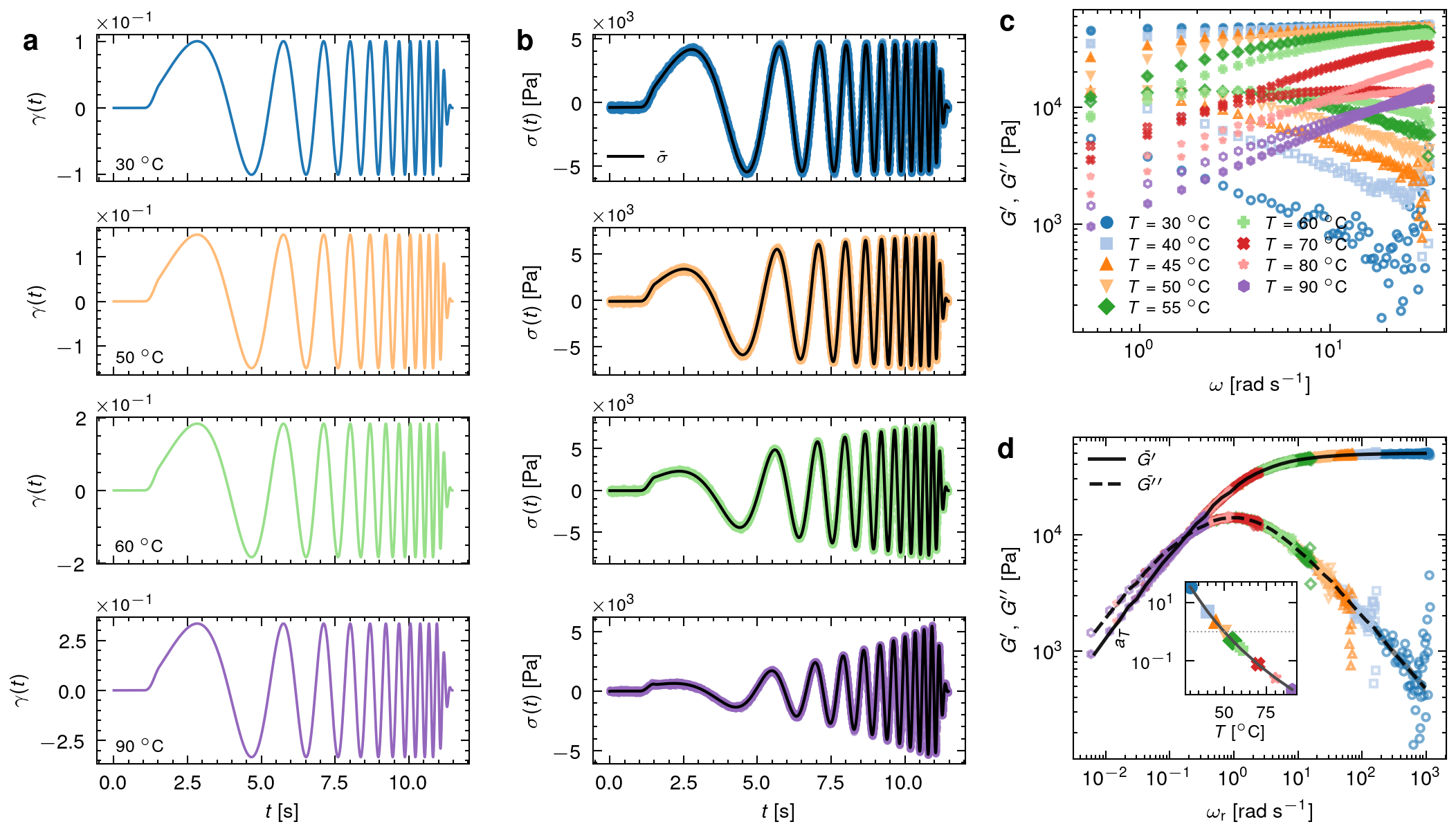}
    \caption{Synthetic FMM benchmark at 3\% additive stress noise. The figure follows the same layout as Fig.~2 of the main text: imposed chirp strains, time-domain stress fits with posterior uncertainty, unshifted frequency-domain spectra, shifted reduced-frequency master curve, and inferred horizontal shift factors.}
    \label{fig:si_fmm_noise003}
\end{figure*}

Figure~\ref{fig:si_fmm_noise003} shows the joint fit for the 3\% noise case. At this lowest noise level, the stress trajectories are reproduced with very small residuals, and the recovered shift factors closely follow the generating WLF curve. The reduced-frequency representation collapses the independently transformed spectra onto the FMM master curve, indicating that the time-domain likelihood resolves both the fractional constitutive parameters and the latent time-rescaling variables with high precision.

\begin{figure*}[t]
    \centering
    \includegraphics[width=1\linewidth]{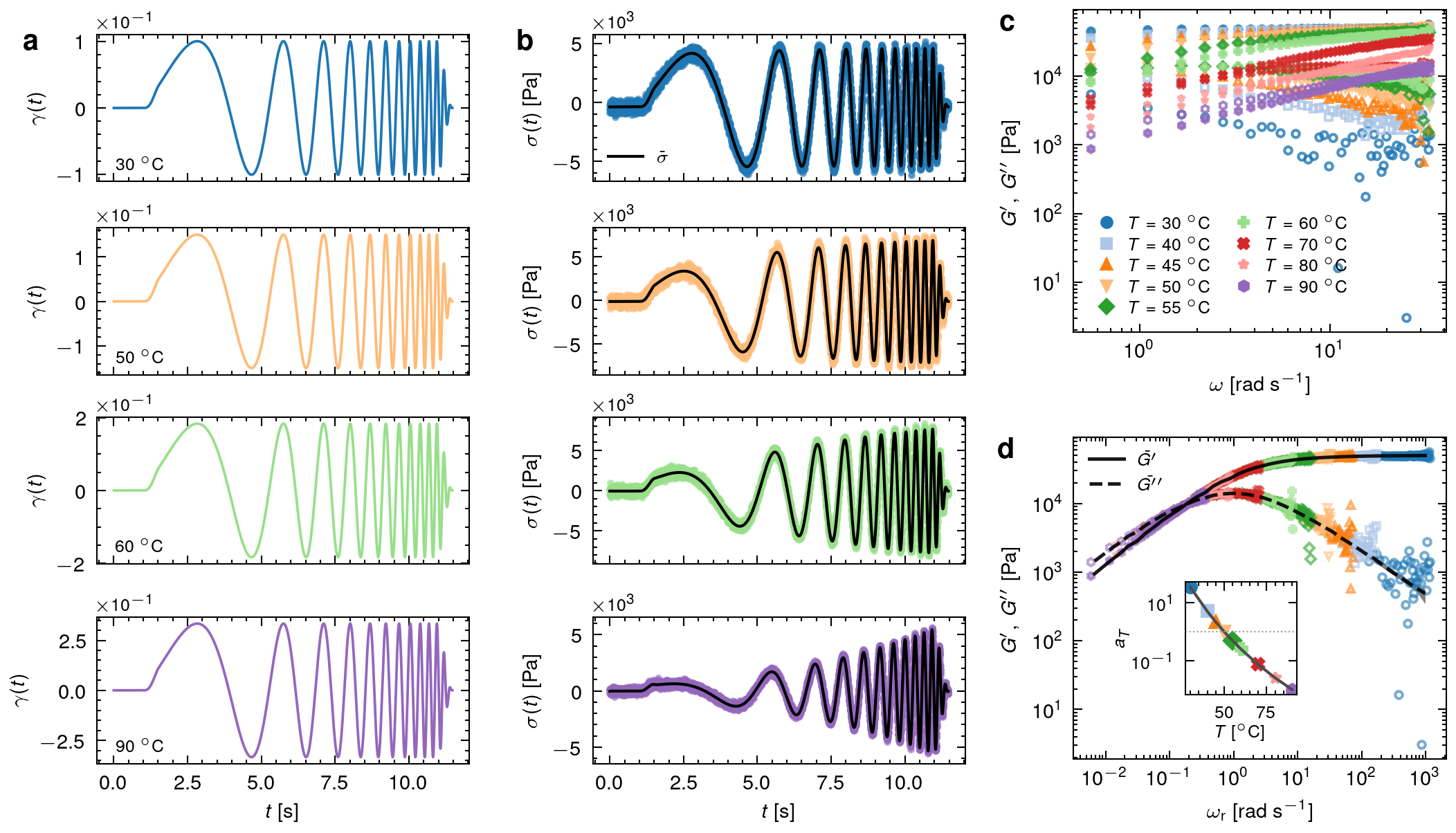}
    \caption{Synthetic FMM benchmark at 8\% additive stress noise. The figure follows the same layout as Fig.~2 of the main text.}
    \label{fig:si_fmm_noise008}
\end{figure*}

Figure~\ref{fig:si_fmm_noise008} shows the corresponding result for 8\% stress noise. The time-domain predictions remain stable across the full temperature range, despite the visibly increased scatter in the stress response and the frequency-domain moduli obtained from separate DFT reductions. The inferred shifts continue to collapse the spectra onto a single reduced-frequency master curve, indicating that the pooled likelihood leverages the shared constitutive structure to regularize inference across temperatures.

\begin{figure*}[!t]
    \centering
    \includegraphics[width=1\linewidth]{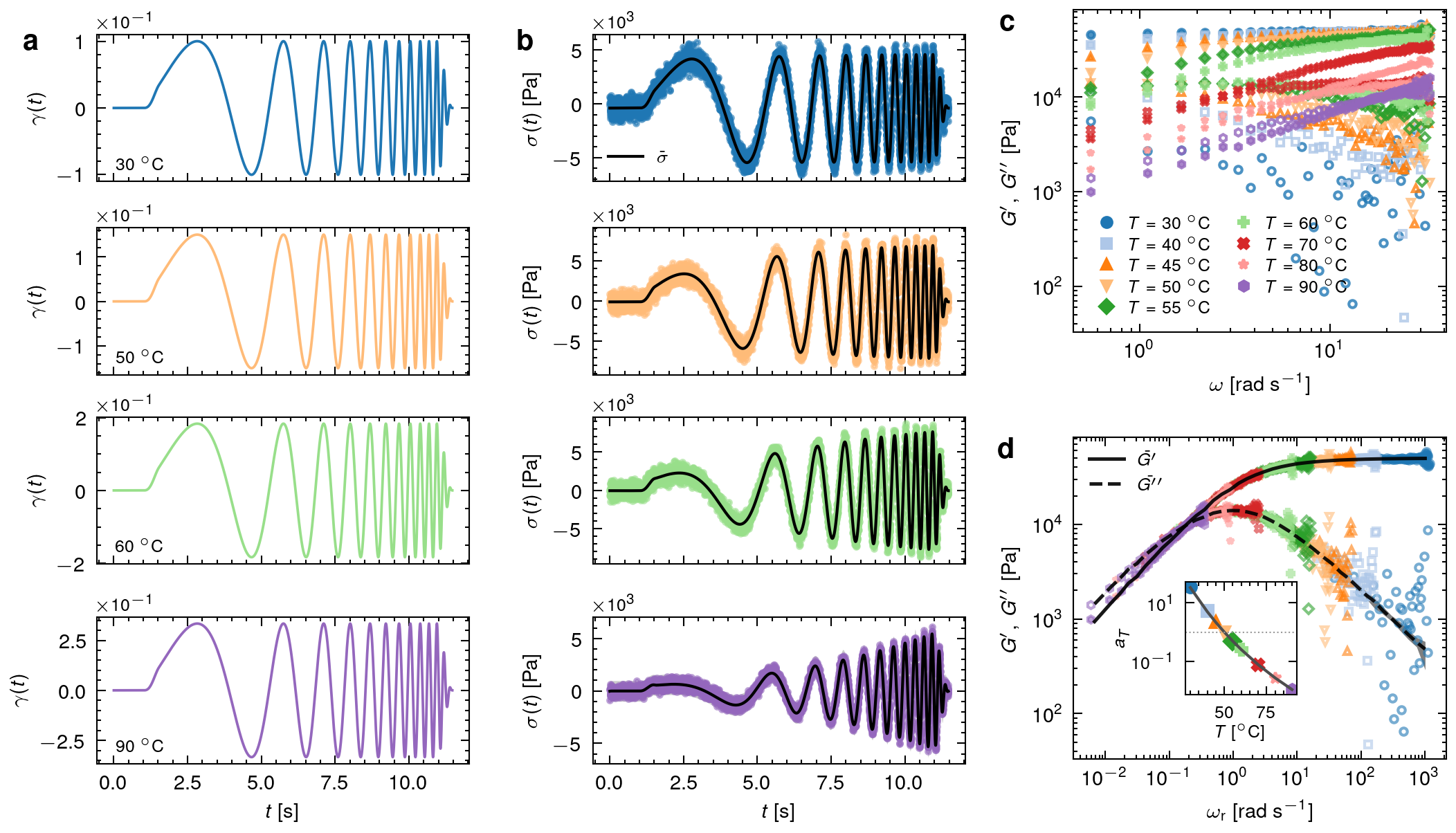}
    \caption{Synthetic FMM benchmark at 15\% additive stress noise. The figure follows the same layout as Fig.~2 of the main text.}
    \label{fig:si_fmm_noise015}
\end{figure*}

Figure~\ref{fig:si_fmm_noise015} reports the 15\% noise case. Although the direct spectral estimates become substantially noisier, particularly for the loss modulus, the joint time-domain fit still recovers the dominant fractional-relaxation behavior. The collapse of the shifted spectra remains clear, illustrating that the inference is not driven by fitting each noisy spectrum independently but by estimating a shared memory law from all raw chirp records simultaneously.

\begin{figure*}[t]
    \centering
    \includegraphics[width=1\linewidth]{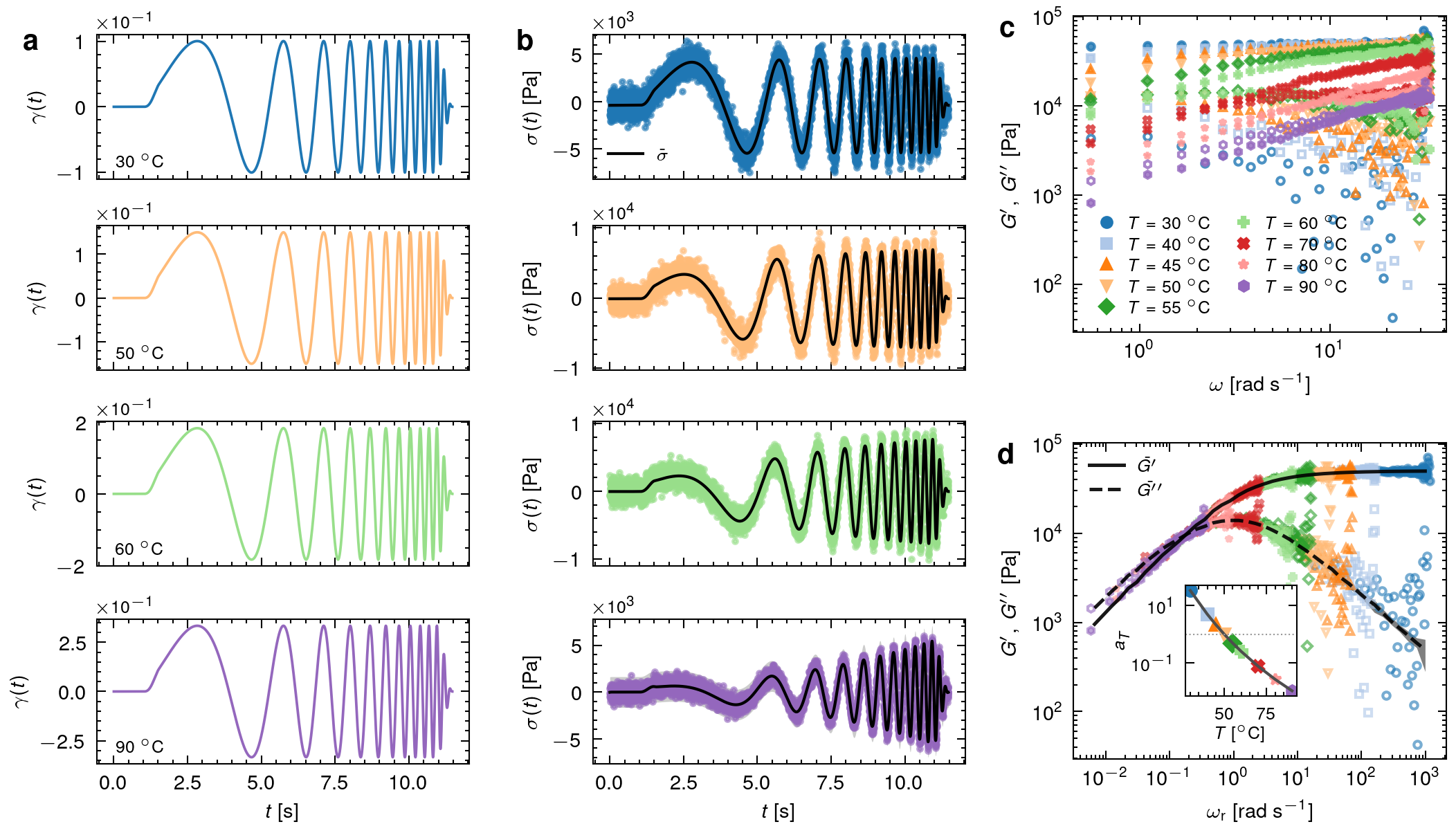}
    \caption{Synthetic FMM benchmark at 25\% additive stress noise. The figure follows the same layout as Fig.~2 of the main text.}
    \label{fig:si_fmm_noise025}
\end{figure*}

Figure~\ref{fig:si_fmm_noise025} gives the highest-noise case, with 25\% additive stress noise. This case provides a stringent test of the latent-variable formulation. Even when the individual DFT-derived spectra are highly scattered, the joint model retains the correct qualitative master-curve structure and estimates shift factors consistent with the underlying trend. The increased uncertainty at this noise level is reflected in the broader posterior bands and in the reduced ability to distinguish nearly nested model variants, as discussed in the BIC analysis in the main text.

Across the four additional noise levels, the same qualitative conclusions hold as in the 5\% case presented in the main text. The pooled time-domain likelihood recovers the shared fractional constitutive law, estimates physically meaningful horizontal shift factors without imposing a WLF form, and produces a reduced-frequency master curve consistent with the generating model. Increasing the stress noise mainly broadens the posterior uncertainty and degrades the independent DFT estimates, rather than causing a systematic failure of the time-domain inference. These results support the robustness of the proposed latent time-rescaling formulation over a broad range of experimentally relevant noise amplitudes.

\begin{figure*}[t]
    \centering
    \includegraphics[width=1\linewidth]{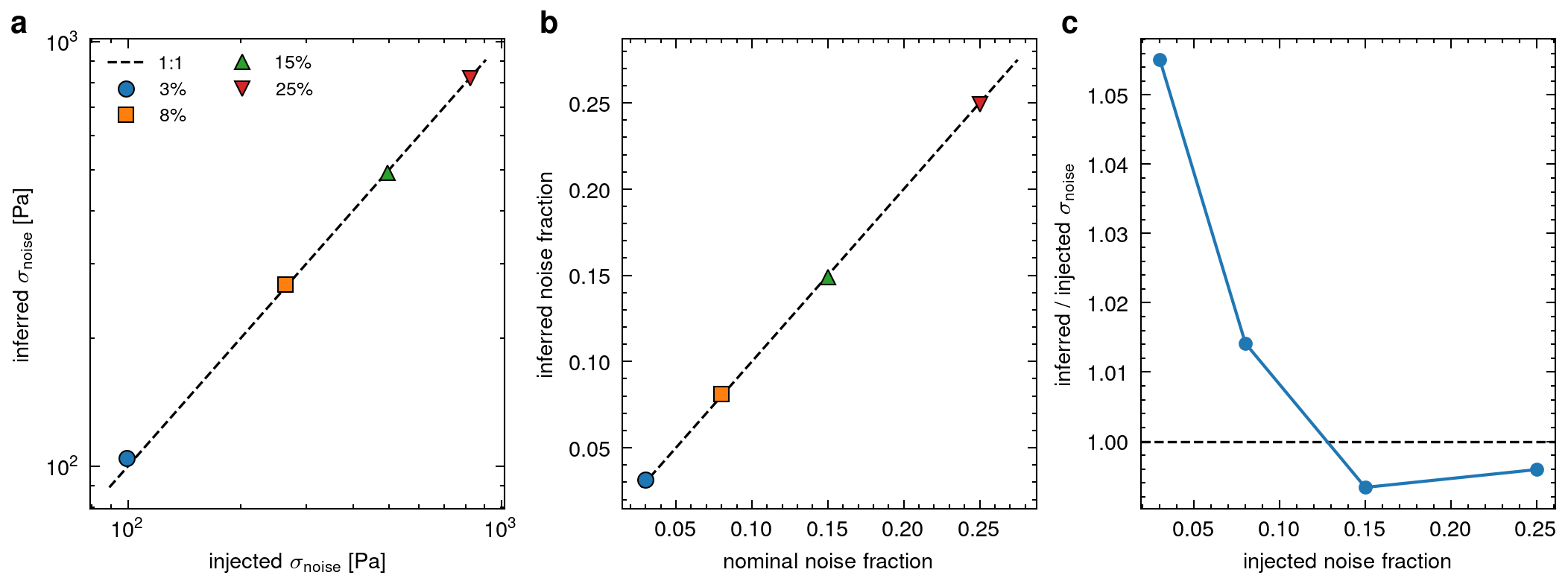}
    \caption{Recovery of the observational noise level in the synthetic FMM noise benchmark. \textbf{a} GP-inferred homoscedastic noise scale plotted against the injected stress-noise scale in physical units of Pa; the identity line indicates perfect recovery. \textbf{b} The same comparison expressed as a percentage of the noiseless stress scale. \textbf{c} Ratio of inferred to injected noise as a function of injected noise level, showing that the learned Gaussian likelihood accurately tracks the imposed homoscedastic noise across the tested range.}
    \label{fig:si_fmm_noise_inferred_vs_real}
\end{figure*}

Figure~\ref{fig:si_fmm_noise_inferred_vs_real} summarizes the recovery of the Gaussian likelihood noise scale across the synthetic FMM noise benchmarks. The inferred homoscedastic noise closely follows the injected noise level, both in absolute stress units and when expressed relative to the noiseless stress scale, indicating that the GP model correctly identifies the observational noise rather than absorbing it into the constitutive response.

%% ------------------------------------------------------------
\subsection*{S2.3.\; Shift-factor comparison with \texttt{mastercurves}}
\label{sec:si_fmm_mastercurves}
%% ------------------------------------------------------------

As an external comparison for the inferred horizontal shifts, we also compared the latent shift factors obtained by the present time-domain method with shift factors obtained using the \texttt{mastercurves} package of \citet{lennon_mckinley_swan_2023}. This comparison tests whether the probabilistic time-domain inference recovers shifts consistent with a well-established method for frequency-domain master-curve construction.

\begin{figure*}[t]
    \centering
    \includegraphics[width=1\linewidth]{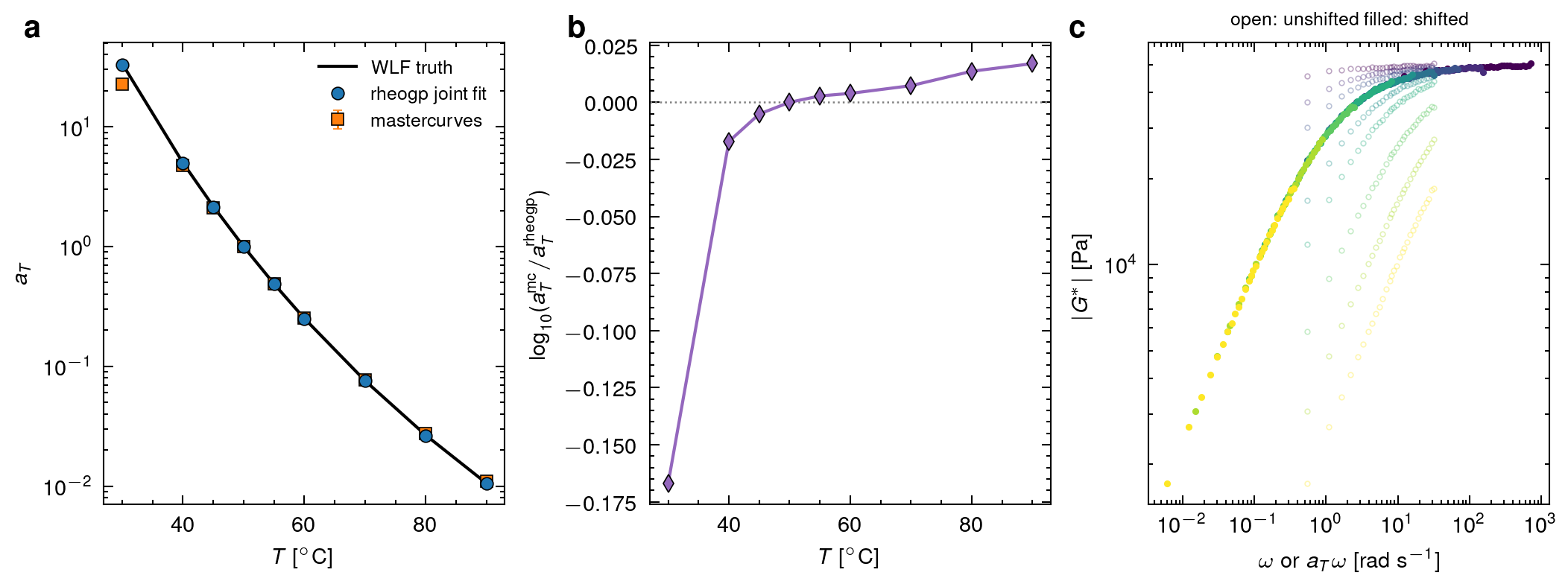}
    \caption{Comparison of inferred horizontal shift factors with the \texttt{mastercurves} package. \textbf{(a)} Horizontal shift factor $a_T$ as a function of temperature for the present time-domain GP inference and for the \texttt{mastercurves} construction. \textbf{(b)} Ratio between the logarithmic shift factors obtained by the present method and by \texttt{mastercurves}. \textbf{(c)} Master curve obtained using \texttt{mastercurves}. Agreement between the inferred shifts supports the interpretation of the latent time-rescaling variables as physically meaningful TTS shift factors.}
    \label{fig:si_mc_fmm_shift_factors}
\end{figure*}

Figure~\ref{fig:si_mc_fmm_shift_factors} shows the resulting comparison. Panel~(a) compares $a_T$ as a function of temperature for the present method and for the \texttt{mastercurves} result. Panel~(b) reports the ratio between the logarithmic shifts obtained by the two methods, highlighting deviations in the inferred horizontal rescaling. Panel~(c) shows the master curve obtained using \texttt{mastercurves}. The two methods give nearly identical shift factors over most of the temperature range, with only a small discrepancy at the lowest temperature. This close agreement supports the interpretation of the latent variables inferred by the GP model as physically meaningful horizontal shift factors.

%% ------------------------------------------------------------
\subsection*{S2.4.\; Prediction of a broadband test chirp}
\label{sec:si_fmm_broadband}
%% ------------------------------------------------------------

\begin{figure*}[t]
    \centering
    \includegraphics[width=1\linewidth]{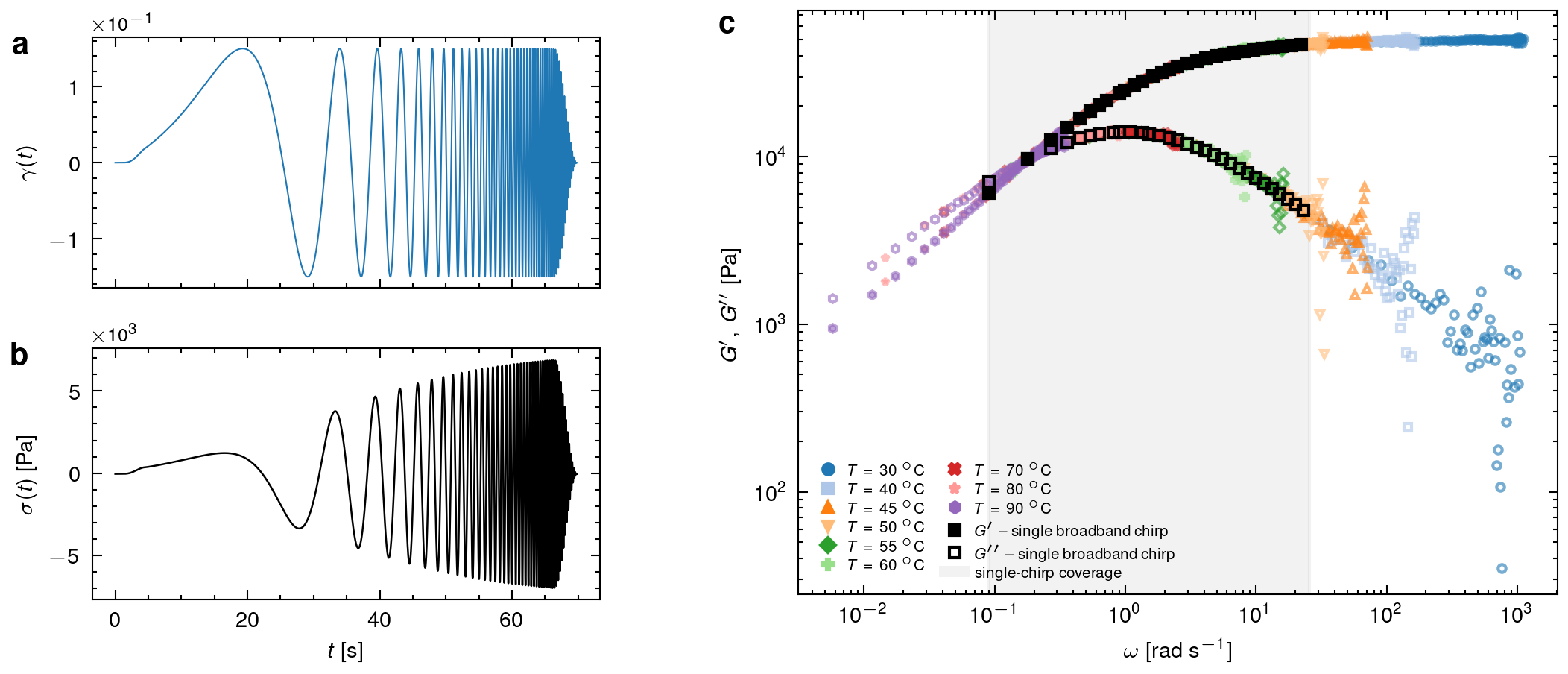}
    \caption{\textbf{(a)} Unseen chirp and \textbf{(b)} the prediction of the stress response using the physics-informed features computed from the constitutive parameters inferred in the joint multi-temperature fit. \textbf{(c)} Discrete Fourier transform of the predicted stress compared to the master curve from Fig.~2 in the main manuscript.}
    \label{fig:si_fmm_broadband_demo}
\end{figure*}

Figure~\ref{fig:si_fmm_broadband_demo} shows the stress response predicted by the trained GP for the broadband test chirp discussed in the main text, which is not included in the training data.

%% TODO: the caption below was a copy of the mastercurves caption
%%       (Fig. S9) in the previous draft; replace with the final caption.

%% ------------------------------------------------------------
\subsection*{S2.5.\; Effective sample size and BIC model selection}
\label{sec:si_fmm_model_selection}
%% ------------------------------------------------------------

We then investigated how the treatment of sample size affects Bayesian information criterion (BIC) model selection across noise levels. Since chirp time series contain many closely spaced samples, the nominal number of time points, $N$, can overstate the amount of independent information when residuals are temporally correlated. We therefore compared three BIC variants: one using the full number of samples in both the likelihood and penalty terms, one using the full likelihood but replacing the penalty term by an effective sample size, $N_{\mathrm{eff}}$, and one in which both the likelihood contribution and the penalty term are evaluated using $N_{\mathrm{eff}}$.

\begin{figure*}[t]
    \centering
    \includegraphics[width=1\linewidth]{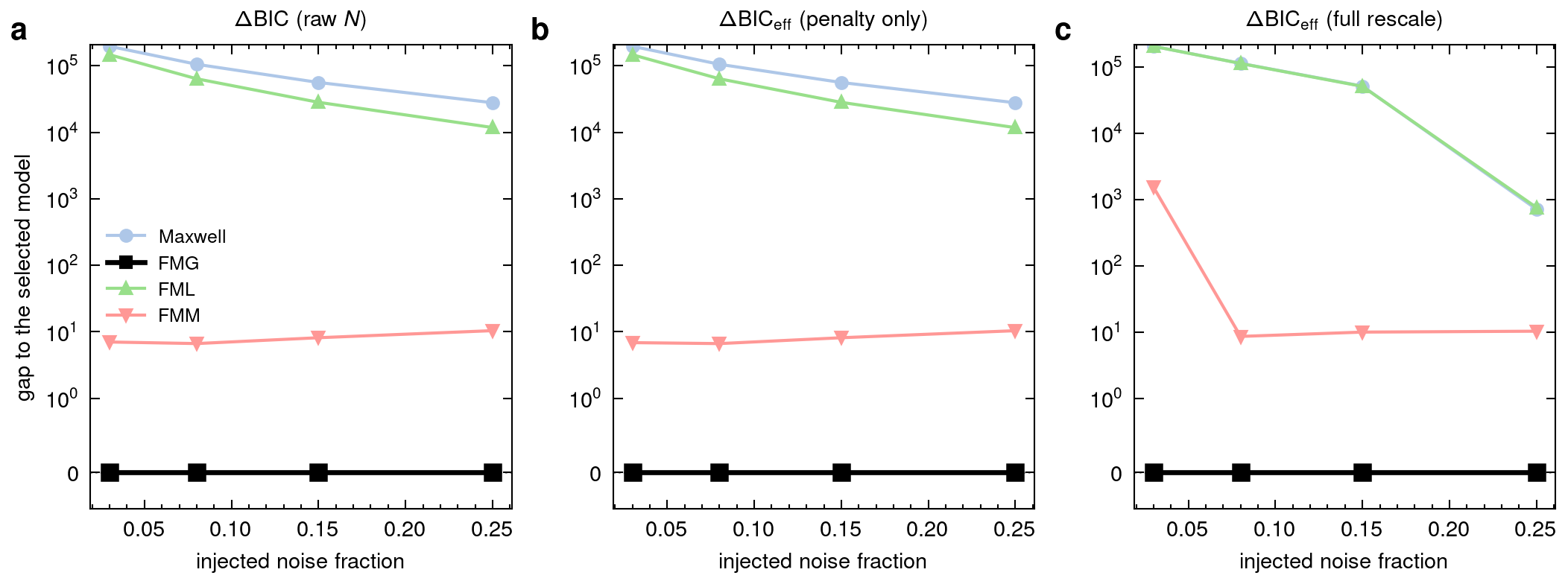}
    \caption{Effect of effective sample size on BIC model selection across synthetic FMM noise levels. \textbf{(a)} $\Delta\mathrm{BIC}$ scores computed using the full number of time-domain samples, $N$, in both the likelihood and penalty terms. \textbf{(b)} $\Delta\mathrm{BIC}$ scores computed using the full likelihood while replacing the penalty sample size by the effective sample size, $N_{\mathrm{eff}}$. \textbf{(c)} Criterion obtained when both the likelihood and penalty terms are evaluated using $N_{\mathrm{eff}}$. The panels report the model-selection decision as a function of injected noise level, showing that the selected model is robust to the effective-sample-size correction.}
    \label{fig:si_fmm_noise_model_selection}
\end{figure*}

Figure~\ref{fig:si_fmm_noise_model_selection} summarizes the resulting model-selection decisions as a function of injected noise level. Panel~(a) shows $\Delta\mathrm{BIC}$ scores computed using the full number of samples, $N$. Panel~(b) shows the corresponding scores when only the complexity penalty is computed using $N_{\mathrm{eff}}$. Panel~(c) shows the criterion obtained when both the likelihood and penalty terms are scaled by $N_{\mathrm{eff}}$. Across these variants, the selected model remains unchanged for the tested synthetic-noise cases, indicating that the model-selection outcome is robust to the effective-sample-size correction. This insensitivity is expected for the synthetic benchmark because the injected observational noise is independent and approximately white. Consistent with this, Fig.~\ref{fig:si_fmm_residual_autocorr} shows that the integrated residual autocorrelation time remains close to its minimum value, with most temperature records corresponding to roughly one effectively independent sample per time step. Thus, the effective sample size is nearly equal to the nominal number of samples, and replacing $N$ by $N_{\mathrm{eff}}$ produces only a small correction to the BIC penalty.

%% ============================================================
\section*{S3.\; Complementary results for the DGEBA--IPD dataset}
\label{sec:si_dgeba_ipd_complementary}
%% ============================================================

We next applied the joint time-domain inference framework to the experimental DGEBA--IPD dataset. In contrast to the synthetic FMM benchmark, the experimental data have no prescribed constitutive parameters, known horizontal shift factors, or injected noise levels. We therefore assess the inference through optimization convergence, the fitted noise model, time-domain residuals, residual autocorrelation, constitutive-model selection, and comparison with an independent frequency-domain master-curve construction. Together, these diagnostics test whether the inferred constitutive response and latent horizontal shifts provide a stable and internally consistent description of the multi-temperature chirp measurements.

\begin{figure*}[t]
    \centering
    \includegraphics[width=\linewidth]{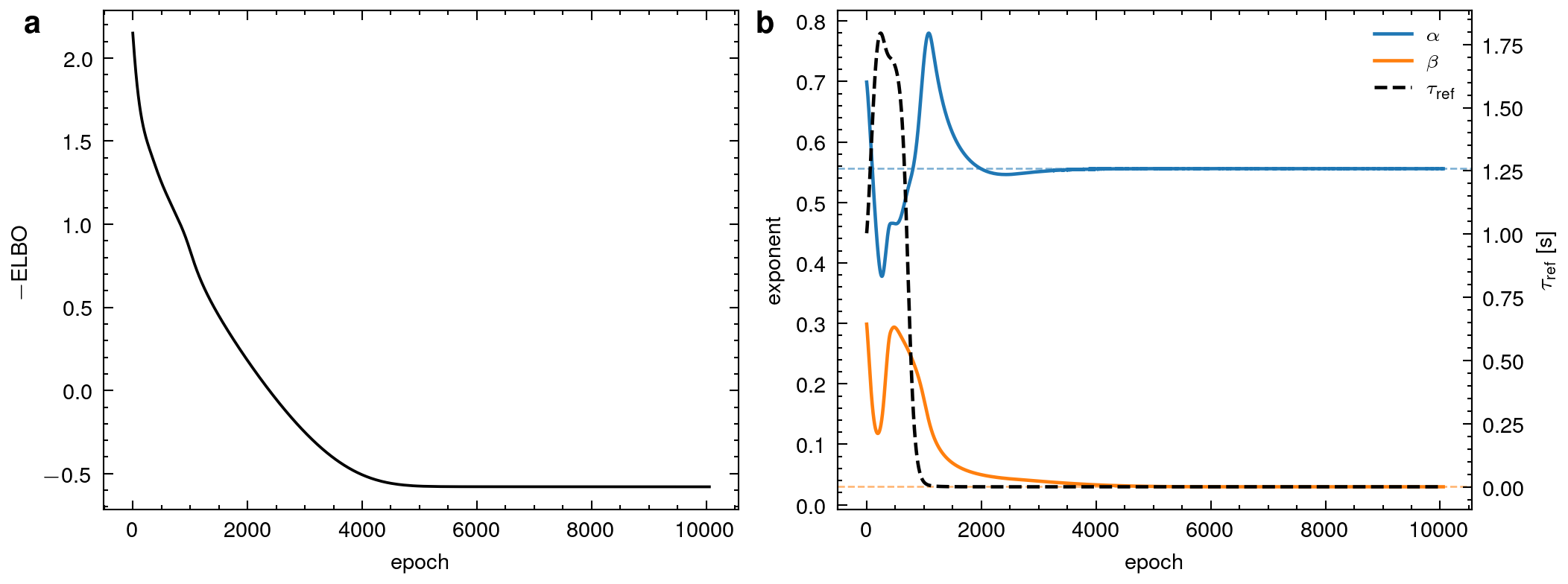}
    \caption{Convergence diagnostics for the experimental DGEBA--IPD dataset. The optimization histories of \textbf{(a)} the ELBO and \textbf{(b)} model parameters are used to assess the stability of the joint time-domain inference and to verify that the inferred constitutive response and horizontal shift factors correspond to a converged solution.}
    \label{fig:si_dgeba_ipd_convergence}
\end{figure*}

Figure~\ref{fig:si_dgeba_ipd_convergence} shows the optimization history for the DGEBA--IPD inference. The convergence traces provide a direct check that the variational objective and inferred quantities approach stable values before the fitted response and latent shift factors are interpreted. Establishing convergence is particularly important for the experimental dataset because model adequacy cannot be evaluated by comparison with known generating parameters.

\begin{figure*}[t]
    \centering
    \includegraphics[width=\linewidth]{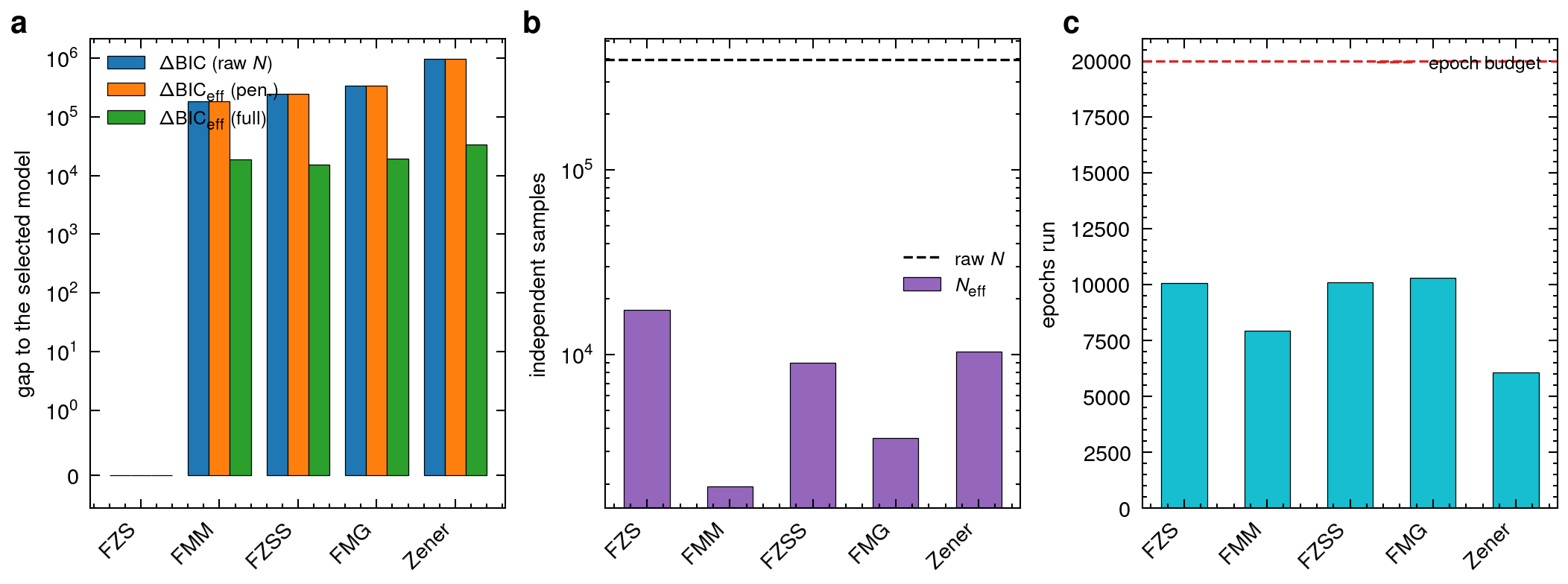}
    \caption{\textbf{(a)} BIC model selection for the experimental DGEBA--IPD dataset, computed using the nominal sample size $N$ and the effective sample size $N_{\mathrm{eff}}$. \textbf{(b)} Independent samples resulting from each fit. \textbf{(c)} Epochs used for each fit and its budget.}
    \label{fig:si_dgeba_ipd_model_selection}
\end{figure*}

We next compared the candidate constitutive models using the Bayesian information criterion. Figure~\ref{fig:si_dgeba_ipd_model_selection} summarizes the model-selection results for the experimental data. Because the residuals can be temporally correlated, the comparison includes the effect of replacing the nominal number of time-domain samples, $N$, with the effective sample size, $N_{\mathrm{eff}}$, in the information criterion. Reporting BIC differences relative to the best-supported candidate makes it possible to assess both the preferred constitutive description and the sensitivity of the ranking to the treatment of temporal correlation.

%% TODO: this figure environment was missing in the previous draft
%%       (the label was referenced but never defined). Check the file
%%       name and replace the placeholder caption.

\begin{figure*}[t]
    \centering
    \includegraphics[width=\linewidth]{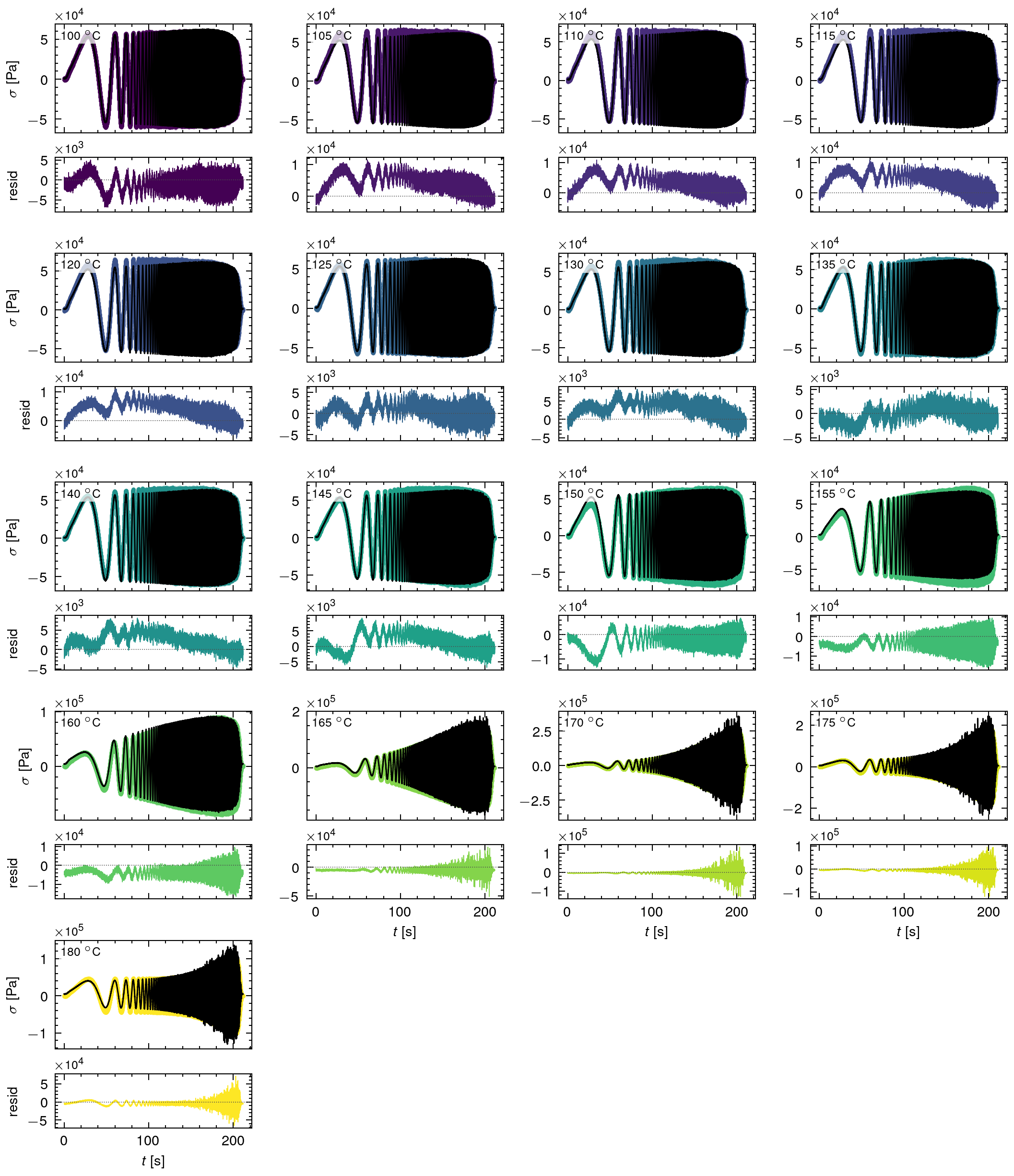}
    \caption{Time-domain residuals for the experimental DGEBA--IPD chirp records. Residuals are shown separately for each temperature-dependent record and provide a diagnostic of measurement variability, instrumental effects, and possible constitutive-model mismatch after joint time-domain inference.}
    \label{fig:si_dgeba_ipd_residuals_per_chirp}
\end{figure*}

Figure~\ref{fig:si_dgeba_ipd_residuals_per_chirp} shows the time-domain residuals for the individual DGEBA--IPD chirp records. These residuals provide a record-by-record diagnostic of the fitted constitutive response. In contrast to the synthetic case, departures from zero can reflect not only observational noise but also instrumental effects, experimental variability, preprocessing, and deviations of the material response from the selected constitutive model. Their temporal and temperature-dependent structure therefore reveals where systematic discrepancies remain after joint inference.

\begin{figure*}[t]
    \centering
    \includegraphics[width=\linewidth]{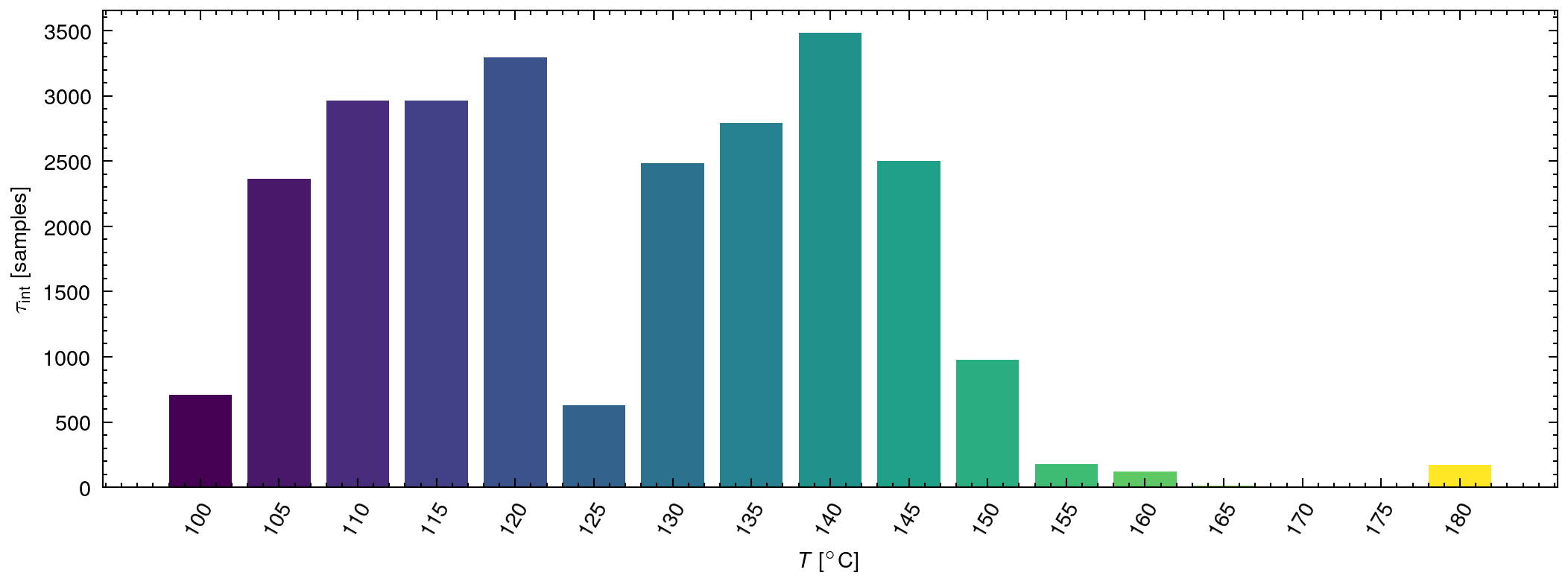}
    \caption{Residual autocorrelation analysis for the experimental DGEBA--IPD dataset. The integrated autocorrelation time, $\tau_{\mathrm{int}}$, is reported for each temperature-dependent chirp record. These values quantify temporal dependence in the residuals and determine the corresponding effective number of independent observations used in the model-selection analysis.}
    \label{fig:si_dgeba_ipd_residual_autocorr}
\end{figure*}

To quantify temporal dependence in the experimental residuals, we computed the integrated autocorrelation time for each chirp record. Figure~\ref{fig:si_dgeba_ipd_residual_autocorr} reports the resulting autocorrelation time scale, $\tau_{\mathrm{int}}$, as a function of temperature. Temporal correlation reduces the amount of independent information contained in the densely sampled time series; accordingly, these estimates provide the basis for defining an effective sample size, $N_{\mathrm{eff}}$, for model-selection diagnostics. Differences among temperatures can reflect changes in measurement conditions, instrumental response, preprocessing, or unmodeled material behavior.

\begin{figure*}[t]
    \centering
    \includegraphics[width=\linewidth]{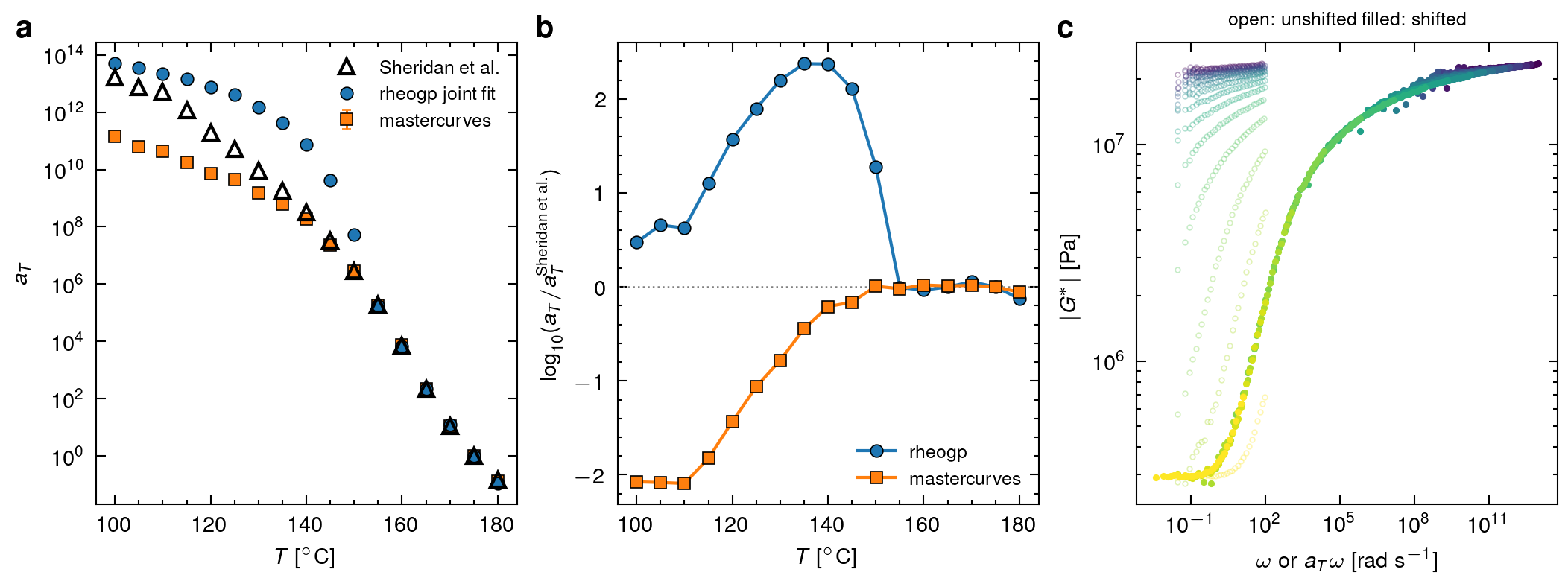}
    \caption{\textbf{(a)} Comparison of horizontal shift factors for the experimental DGEBA--IPD dataset. The temperature-dependent shift factors inferred jointly by the time-domain GP framework are compared with those obtained using the \texttt{mastercurves} package. \textbf{(b)} Agreement between the two estimates supports the interpretation of the latent time-rescaling variables as physically meaningful time--temperature superposition factors, while deviations identify temperatures at which the inferred rescaling is method-dependent. \textbf{(c)} Master curve obtained using the \texttt{mastercurves} package.}
    \label{fig:si_dgeba_ipd_mc_shift_factors}
\end{figure*}

As an external assessment of the latent time-rescaling variables, we compared the horizontal shift factors inferred by the time-domain GP framework with those obtained using the \texttt{mastercurves} package of \citet{lennon_mckinley_swan_2023}. Figure~\ref{fig:si_dgeba_ipd_mc_shift_factors} presents the temperature-dependent shifts obtained by the two methods. The GP estimates are inferred jointly from the raw stress--strain records and the shared constitutive response, whereas the \texttt{mastercurves} estimates are obtained from a frequency-domain superposition procedure. Their comparison therefore tests whether two distinct inference strategies identify a consistent temperature dependence of the material time scale.

\begin{figure*}[t]
    \centering
    \includegraphics[width=\linewidth]{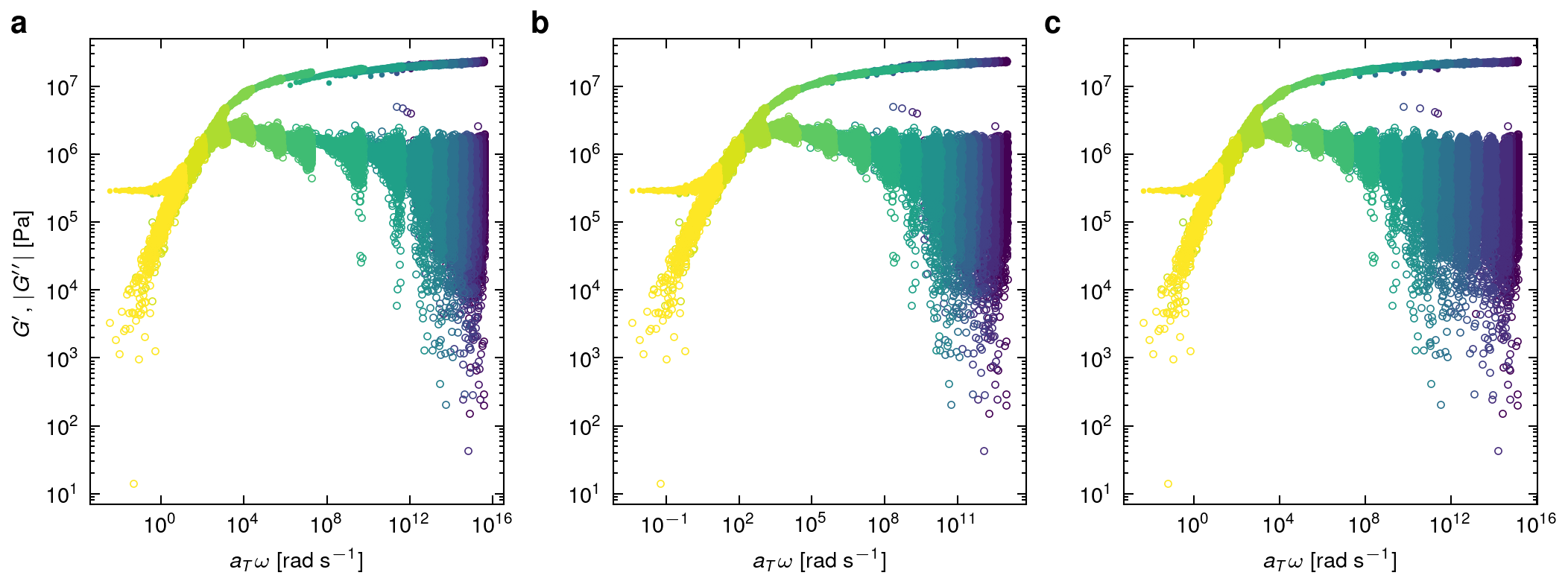}
    \caption{Frequency-domain master curve for the experimental DGEBA--IPD dataset constructed using the approach from \textbf{(a)} this work, \textbf{(b)} the \texttt{mastercurves} package and \textbf{(c)} the shifts from \citet{sheridan_zauscher_brinson_2024}. The shifted storage and loss moduli are plotted against reduced frequency to assess the quality of time--temperature superposition and to provide an independent comparison with the response inferred from the pooled time-domain chirp records.}
    \label{fig:si_dgeba_ipd_mc_master_curve}
\end{figure*}

Figure~\ref{fig:si_dgeba_ipd_mc_master_curve} shows the corresponding frequency-domain master curve constructed using \texttt{mastercurves}. The storage and loss moduli are plotted against reduced frequency after applying the independently estimated horizontal shifts. The degree of overlap among the temperature-dependent records provides an external assessment of time--temperature superposition and complements the shift-factor comparison in Fig.~\ref{fig:si_dgeba_ipd_mc_shift_factors}. It also provides an independent frequency-domain representation against which the broadband response inferred directly from the time-domain chirps can be evaluated.

Overall, the DGEBA--IPD results provide an experimental counterpart to the synthetic FMM benchmark. Because the true constitutive parameters, horizontal shifts, and observational-noise levels are unknown, validation relies on mutually consistent diagnostic evidence rather than recovery of prescribed generating quantities. The convergence history establishes the stability of the optimization, the noise and residual analyses characterize the unexplained experimental variability, and the autocorrelation analysis quantifies the reduction in effective sample size. The model-selection results identify the constitutive description supported by the pooled time-domain records, while the independent \texttt{mastercurves} comparison assesses the physical consistency of the inferred horizontal shifts and reduced-frequency response. Together, these results demonstrate how the joint time-domain framework can be evaluated when applied to experimental multi-temperature chirp rheometry.

%% ============================================================
\section*{S4.\; Implementation details}
\label{sec:SI_impl}
%% ============================================================

The framework is implemented in Python using GPyTorch~\cite{gardner2018gpytorch} for GP inference and PyTorch for automatic differentiation and gradient-based optimization. All experiments were run on a single NVIDIA H200 MIG GPU with $71~\mathrm{GB}$ of VRAM. The learning rates were fixed at 0.001 for the GP hyperparameters and 0.005 for the constitutive parameters. Optimization was performed for a maximum of $20\,000$ epochs, with early stopping triggered when the ELBO did not improve for 500 consecutive epochs. Two inducing points were used for models with one input feature, such as the fractional Maxwell model, whereas three were used for models with two input features, such as the fractional Zener model.

Code and data supporting this study are openly available at \url{https://github.com/mirandi1/rheogp}.

%% ============================================================

%% ============================================================
\section*{S5.\; Applications to additional experimental datasets}
\label{sec:SI_additional_applications}
%% ============================================================

\begin{figure*}[t]
    \centering
    \includegraphics[width=\linewidth]{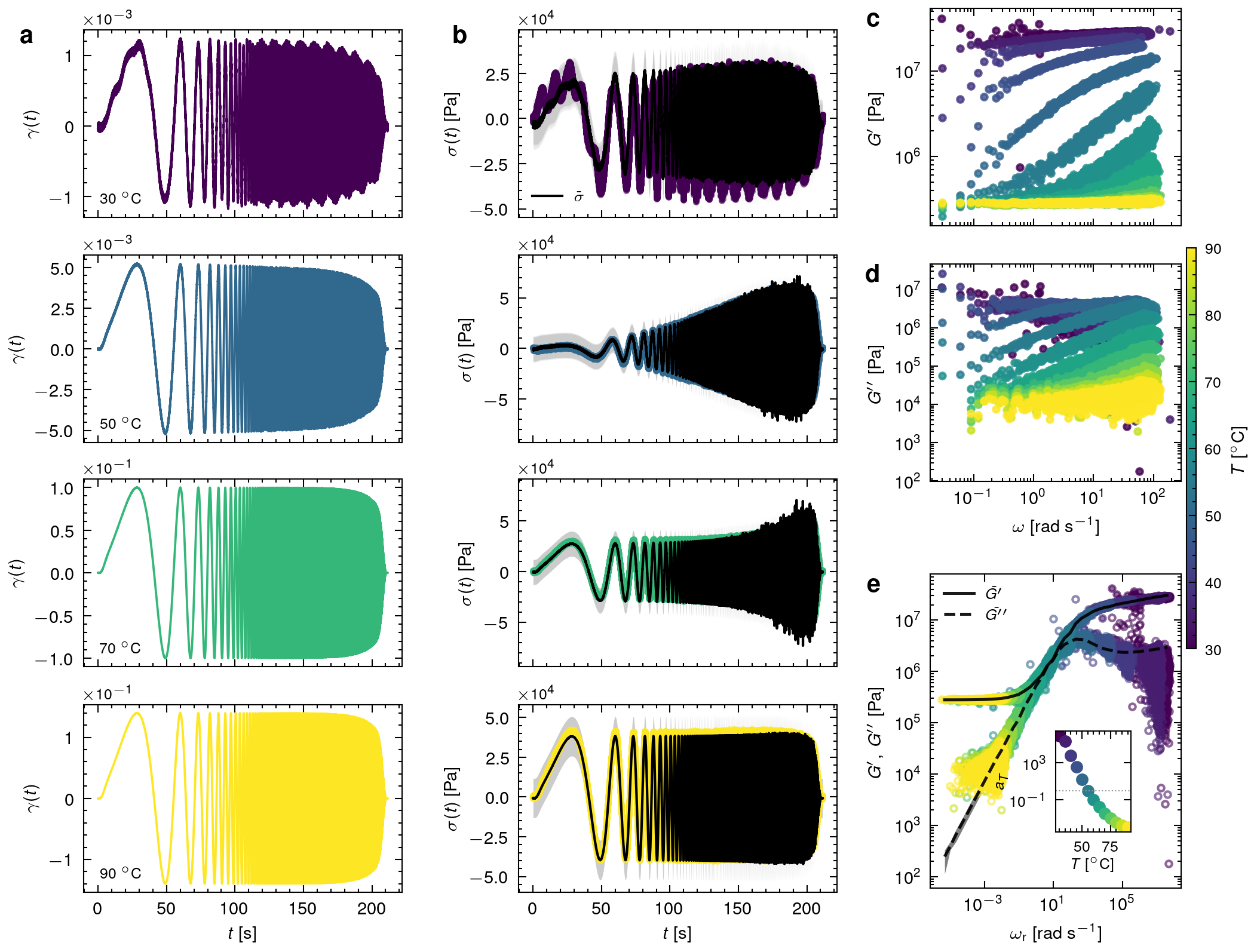}
    \caption{Application of the joint time-domain framework to the photopolymer dataset reported by \citet{sheridan_zauscher_brinson_2024} at $T_{\mathrm{ref}}=55~^{\circ}\mathrm{C}$. Consult the original publication for further details about the material and experimental protocol. The layout follows the same as in Fig.~3 of the main manuscript.}
    \label{fig:si_photopolymer_application}
\end{figure*}

\begin{figure*}[t]
    \centering
    \includegraphics[width=\linewidth]{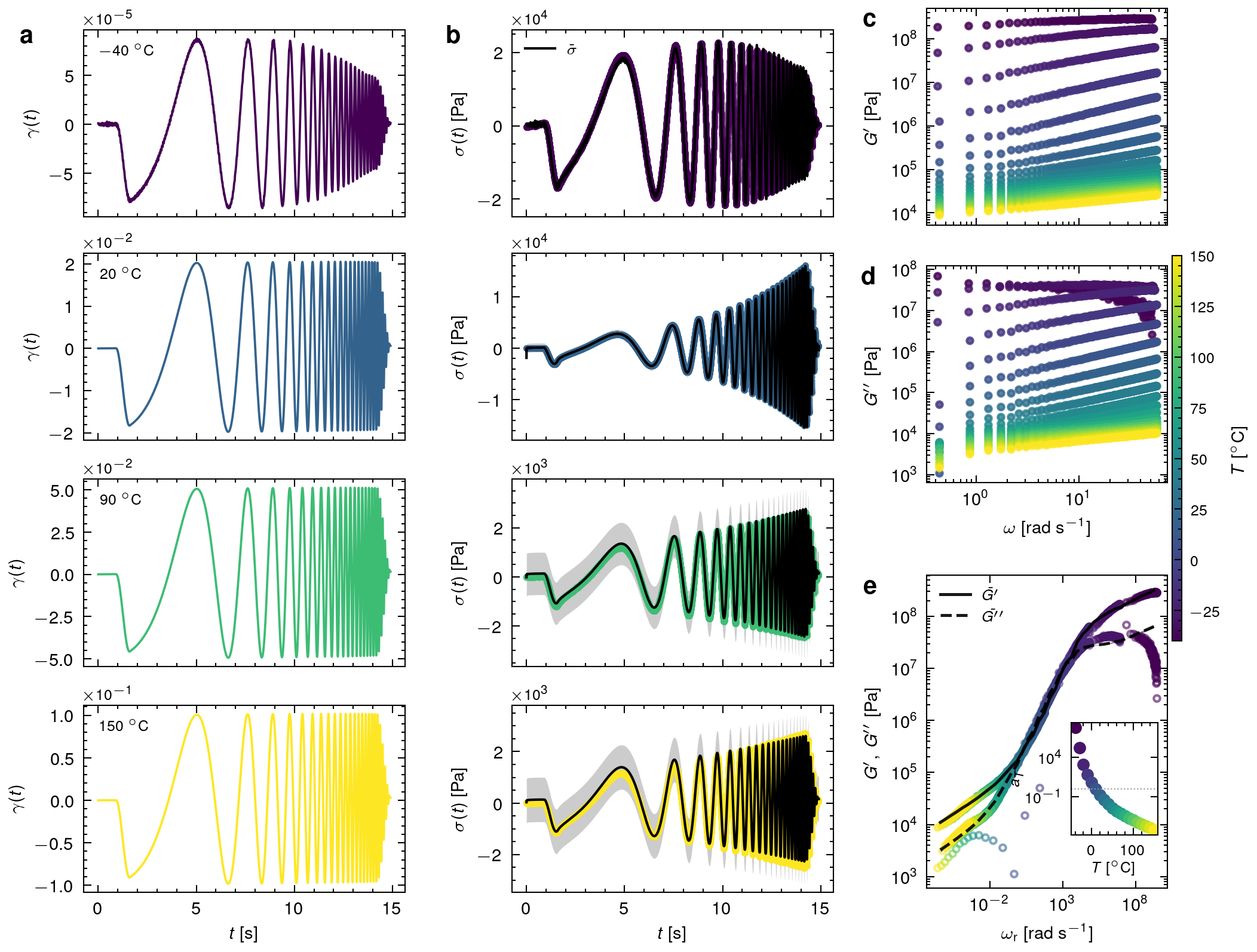}
    \caption{Application of the joint time-domain framework to the acrylic-based adhesive dataset reported by \citet{Perego2026} at $T_{\mathrm{ref}}=10~^{\circ}\mathrm{C}$. Consult the original publication for further details about the material and experimental protocol. The layout follows the same as in Fig.~3 of the main manuscript.}
    \label{fig:si_acrylic_adhesive_application}
\end{figure*}

To assess the applicability of the proposed framework beyond the DGEBA--IPD dataset, we considered two additional experimental systems reported in the literature: the photopolymer dataset of \citet{sheridan_zauscher_brinson_2024} and the acrylic-based adhesive dataset of \citet{Perego2026}. Figure~\ref{fig:si_photopolymer_application} shows the application of the proposed approach to the photopolymer dataset using a fractional Zener model composed of two springpots and a spring. The fitted parameters were $\alpha=0.6364$, $\beta=0.06085$, $\tau_{\mathrm{ref}}=8.046\times10^{-3}$ s, $G_c=1.42608\times10^{7}$ Pa with a 95\% confidence interval of $[1.41937\times10^{7},\,1.43167\times10^{7}]$, and $G_0=2.78003\times10^{5}$ Pa with a 95\% confidence interval of $[2.76467\times10^{5},\,2.80230\times10^{5}]$. The fit yielded $\mathrm{RMSE}=6172$ Pa, $R^2=0.9485$, and $\mathrm{BIC}=-34216.0$. Figure~\ref{fig:si_acrylic_adhesive_application} presents the corresponding analysis of the acrylic-based adhesive dataset, which accompanies \texttt{hermes-rheo}, an open-source Python package for rheological data analysis. For this material, we used a fractional Zener model composed of three springpots, with the third fractional order denoted by $\nu$. The fitted parameters were $\alpha=0.6406$, $\beta=0.1262$, $\nu=0.1808$, $\tau_{\mathrm{ref}}=1.00015\times10^{-4}~\mathrm{s}$, $G_c=6.7376\times10^{7}~\mathrm{Pa}$ with a 95\% confidence interval of $[6.72536\times10^{7},\,6.76024\times10^{7}]~\mathrm{Pa}$, and $\mathbb{K}=1.05711\times10^{5}~\mathrm{Pa\,s}^{\nu}$ with a 95\% confidence interval of $[1.04194\times10^{5},\,1.06819\times10^{5}]~\mathrm{Pa\,s}^{\nu}$. From $G_c=\mathbb{V}\tau_{\mathrm{ref}}^{-\alpha}$, the corresponding springpot prefactor is $\mathbb{V}=G_c\tau_{\mathrm{ref}}^{\alpha}=1.845\times10^{5}~\mathrm{Pa\,s}^{\alpha}$. The retardation time is therefore $\tau_{\mathrm{ret}}=(\mathbb{V}/\mathbb{K})^{1/(\alpha-\nu)}=3.36~\mathrm{s}$, yielding the equivalent equilibrium modulus $G_e=\mathbb{V}\tau_{\mathrm{ret}}^{-\alpha}=\mathbb{K}\tau_{\mathrm{ret}}^{-\nu}=8.49\times10^{4}~\mathrm{Pa}$. The fit yielded $\mathrm{RMSE}=426.2$, $R^2=0.9914$, and $\mathrm{BIC}=-283056.6$.

\begin{acknowledgments}
See main text.
\end{acknowledgments}

\bibliography{supplementary}